%% file: FRS-journal.tex
\documentclass[11pt]{article}

\usepackage{amsmath}
\usepackage{amssymb}
\usepackage{amsthm}
\usepackage{fullpage}
\usepackage{hyperref}
\usepackage{epsf}
\usepackage{graphicx}
\usepackage{color}

\newtheorem{definition}{Definition}[section]

\newtheorem{theorem}{Theorem}[section]
\newtheorem{lemma}[theorem]{Lemma}

\theoremstyle{definition}
\newtheorem{remark}{Remark}[section]

\newcommand{\p}{\rho}
\newcommand{\F}{\mathbb{F}}

\newcommand{\R}{\mathbb{R}}

\newcommand{\eps}{\varepsilon}
\renewcommand{\epsilon}{\varepsilon}

\newcommand{\eqdef}{\stackrel{{\rm def}}{=}}

\newcommand{\GS}{{\rm GS}}
\newcommand{\low}{{\rm b}}
\newcommand{\high}{{\rm a}}

\renewcommand{\le}{\leqslant}
\renewcommand{\ge}{\geqslant}

\newcommand{\mv}[1]{{\mathbf {#1}}}

\newcommand{\angles}[1]{\langle #1 \rangle}

\renewcommand{\a}{\alpha}
\renewcommand{\b}{\beta}
\newcommand{\g}{\gamma}

\newcommand{\ord}{{\rm ord}_{\F_q}}

\newcommand{\frs}{{\sf FRS}}

\newcommand{\anote}[1]{$\ll$\textsf{#1 --Atri}$\gg$\marginpar{\tiny\bf AR}}

\title{{{\bf Explicit Codes Achieving List Decoding Capacity: \\ 
Error-correction with Optimal Redundancy}}\thanks{A
    preliminary version of this paper~\cite{GR-stoc06} appears in the
    {\em Proceedings of 38th Annual ACM Symposium on Theory of
    Computing} under the title ``Explicit capacity-achieving
    list-decodable codes''.}}

\author{{\sc Venkatesan Guruswami}$^1$\thanks{Research supported by NSF
    Career award CCF-0343672, an Alfred P. Sloan Research Fellowship,
    and a David and Lucile Packard Foundation Fellowship.}
\and 
{\sc Atri Rudra}$^2$\thanks{Research supported by NSF CCF-0343672. This work was 
done when the author was at University of Washington.}}

\date{$^1$ Department of Computer
  Science and Engineering\\
University of Washington \\ Seattle, WA 98195 \\
{\tt venkat@cs.washington.edu}\\
\vspace*{.3cm}
$^2$ Department of Computer
  Science and Engineering\\
State University of New York at Buffalo \\ Buffalo, NY 14260 \\
{\tt atri@cse.buffalo.edu}}

\begin{document}
\maketitle
\begin{abstract}
 We present error-correcting codes that achieve the
information-theoretically best possible trade-off between the rate and
error-correction radius. Specifically, for every $0 < R < 1$ and $\eps
> 0$, we present an explicit construction of error-correcting codes of
rate $R$ that can be list decoded in polynomial time up to a fraction
$(1-R-\eps)$ of {\em worst-case} errors.  At least theoretically, this
meets one of the central challenges in algorithmic coding theory.
  
  Our codes are simple to describe: they are {\em folded Reed-Solomon
    codes}, which are in fact {\em exactly} Reed-Solomon (RS) codes, but
    viewed as a code over a larger alphabet by careful bundling of
    codeword symbols. Given the ubiquity of RS codes, this is an
    appealing feature of our result, and in fact our methods directly
    yield better decoding algorithms for RS codes when errors occur in
    {\em phased bursts}. 

 The alphabet size of these folded RS codes is polynomial in the block
  length. We are able to reduce this to a constant (depending on
  $\eps$) using ideas concerning ``list recovery'' and
  expander-based codes from \cite{GI-focs01,GI-ieeejl}. Concatenating
  the folded RS codes with suitable inner codes also gives us
  polynomial time constructible binary codes that can be efficiently
  list decoded up to the Zyablov bound, i.e., up to twice the radius
  achieved by the standard GMD decoding of concatenated codes.

\end{abstract}

\newpage
\section{Introduction}

\subsection{Background on List Decoding}

Error-correcting codes enable reliable communication of messages over
a noisy channel by cleverly introducing redundancy into the message to
encode it into a codeword, which is then transmitted on the channel.
This is accompanied by a decoding procedure that recovers the correct
message even when several symbols in the transmitted codeword are
corrupted. In this work, we focus on the adversarial or worst-case
model of errors --- we do not assume anything about how the errors and
error locations are distributed beyond an upper bound on the total
number of errors that may be caused. 
The central trade-off in this theory is the one between the amount of
redundancy needed and the fraction of errors that can be corrected.
The redundancy is measured by the {\em rate} of the code, which is the
ratio of the the number of information symbols in the message to that
in the codeword --- thus, for a code with encoding function $E :
\Sigma^k \rightarrow \Sigma^n$, the rate equals $k/n$.  The {\em block
  length} of the code equals $n$, and $\Sigma$ is its {\em alphabet}.

The goal in decoding is to find, given a noisy received word, the
actual codeword that it could have possibly resulted from.  If we
target correcting a fraction $\p$ of errors ($\p$ will be called the
error-correction radius or decoding radius), then this amounts to
finding codewords within (normalized Hamming) distance $\p$ from the
received word. We are guaranteed that there will be a unique such
codeword provided {\em every} two distinct codewords differ on at
least a fraction $2\p$ of positions, or in other words the
relative distance of the code is at least $2\p$.  However, since the
relative distance $\delta$ of a code must satisfy $\delta \le 1-R$
where $R$ is the rate of the code (by the Singleton bound), the best
trade-off between $\p$ and $R$ that unique decoding permits is $\p =
\p_U(R) = (1-R)/2$. But this is an overly pessimistic estimate of the
error-correction radius, since the way Hamming spheres pack in space,
for {\em most} choices of the received word there will be at most one
codeword within distance $\p$ from it even for $\p$ much greater than
$\delta/2$.  Therefore, {\em always} insisting on a unique answer will
preclude decoding most such received words owing to a few pathological
received words that have more than one codeword within distance
roughly $\delta/2$ from them.

A notion called list decoding provides a clean way to get around this
predicament, and yet deal with worst-case error patterns. Under list
decoding, the decoder is required to output a list of all codewords
within distance $\p$ from the received word.  The notion of list
decoding itself is quite old and dates back to work in 1950's by
Elias~\cite{elias} and Wozencraft~\cite{wozencraft}. However, the
algorithmic aspects of list decoding were not revived until the more
recent works \cite{GL,sudan} which studied the problem for
complexity-theoretic motivations.

Let us call a code $C$ {\em $(\p,L)$-list decodable} if the number of
codewords within distance $\p$ of any received word is at most $L$. To
obtain better trade-offs via list decoding, we need $(\p,L)$-list
decodable codes where $L$ is bounded by a polynomial function of the
block length, since this is an {\em a priori} requirement for polynomial
time list decoding.  How large can $\p$ be as a function of $R$ for
which such $(\p,L)$-list decodable codes exist? A standard random
coding argument shows that we can have $\p \ge 1-R - o(1)$ over large
enough alphabets, cf. \cite{ZP,elias91}, and a simple counting
argument shows that $\p$ can be at most $1-R$. Therefore the {\em
list decoding capacity}, i.e., the information-theoretic limit of list
decodability, is given by the trade-off $\p_{\rm cap}(R) = 1-R = 2
\p_U(R)$. Thus list decoding holds the promise of correcting {\em
twice} as many errors as unique decoding, for {\em every} rate.

We note that since the original message ${\cal M}$ has $Rn$ symbols, it is
information-theoretically impossible to perform the decoding if at
most a fraction $(R-\eps)$ of the received symbols agree with the
encoding of ${\cal M}$ (for some $\eps > 0$). This holds even for the erasure
channel, and even if we are told in advance which symbols will be
erased! Therefore, for any given rate, list decoding allows one to
decode up to the largest fraction of errors that one can meaningfully
hope to correct.

\begin{figure}[ptbh]
\begin{center}
\epsfysize=3in
 \epsffile{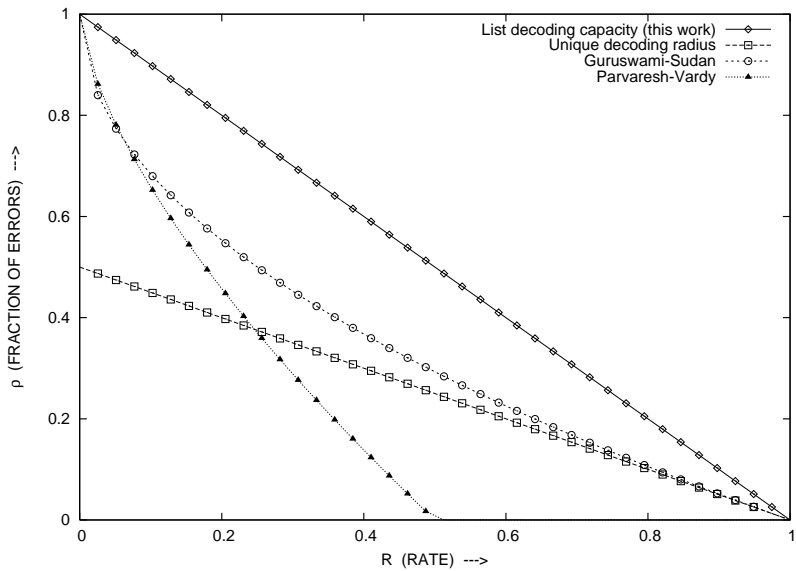}
  \end{center}
  \caption{{\rm Error-correction radius $\p$ plotted against the 
  rate $R$ of the code for known algorithms. The best possible
  trade-off, i.e., capacity, is $\p=1-R$, and our work achieves this.}}
  \label{fig:ld-cap-large}
\end{figure}

The above-mentioned list decodable codes are, however,
non-constructive. In order to realize the potential of list decoding,
one needs explicit constructions of such codes, and on top of that,
polynomial time algorithms to perform list decoding. After essentially
no progress in this direction in over 30 years, the work of
Sudan~\cite{sudan} and improvements to it in \cite{GS98}, achieved
efficient list decoding up to $\p_{\GS}(R) = 1-\sqrt{R}$ errors for
an important family of codes called Reed-Solomon codes. Note that $1
-\sqrt{R} > \p_U(R) = (1-R)/2$ for every rate $R$, $0 < R < 1$, so
this result showed that list decoding can be effectively used to go
beyond the unique decoding radius for every rate (see
Figure~\ref{fig:ld-cap-large}). The ratio $\p_{\GS}(R)/\p_U(R)$
approaches $2$ for rates $R \to 0$, enabling error-correction when the
fraction of errors approaches 100\%, a feature that has found numerous
applications outside coding theory, see for example
\cite{sudan-sigact}, \cite[Chap. 12]{G-thesis}.

Unfortunately, the improvement provided by \cite{GS98} over unique
decoding diminishes for larger rates, which is actually the regime of
greater practical interest. For rates $R \to 1$, the ratio $\frac{\p_{\rm
  GS}(R)}{\p_U(R)}$ approaches $1$, and already for rate $R=1/2$ the
ratio is at most $1.18$. Thus, while the results of \cite{sudan,GS98}
demonstrated that list decoding always, for every rate, enables
correcting more errors than unique decoding, they fell short of
realizing the full quantitative potential of list decoding.

The bound $\p_{\GS}(R)$ stood as the best known error-correction
radius for efficient list decoding for several years. In fact
constructing $(\p,L)$-list decodable codes of rate $R$ for $\p > \p_{\rm
  GS}(R)$ and polynomially bounded $L$, regardless of the complexity
of actually performing list decoding to radius $\p$, itself was
elusive. Some of this difficulty was due to the fact that $1-\sqrt{R}$
is the largest radius for which small list size can be shown
generically, via the so-called Johnson bound to argue about the number
of codewords in Hamming balls using only information on the relative
distance of the code, cf.  \cite{G-johnson}.

In a recent breakthrough paper~\cite{PV-focs05}, Parvaresh and Vardy
presented codes that are list-decodable beyond the $1-\sqrt{R}$ radius
for low rates $R$. The codes they suggest are variants of Reed-Solomon
(RS) codes obtained by evaluating $m \ge 1$ correlated polynomials at
elements of the underlying field (with $m=1$ giving RS codes). For any
$m \ge 1$, they achieve the error-correction radius $\p^{(m)}_{\rm
  PV}(R) = 1 - \sqrt[m+1]{m^m R^m}$. For rates $R \to 0$, choosing $m$
large enough, they can list decode up to radius $1- O(R \log (1/R))$,
which approaches the capacity $1-R$. However, for $R \ge 1/16$, the
best choice of $m$ (the one that maximizes $\p^{(m)}_{\rm PV}(R)$) is
in fact $m=1$, which reverts back to RS codes and the error-correction
radius $1-\sqrt{R}$.  (See Figure~\ref{fig:ld-cap-large} where the
bound $1- \sqrt[3]{4R^2}$ for the case $m=2$ is plotted --- except for
very low rates, it gives a small improvement over $\p_{\GS}(R)$.)
Thus, getting arbitrarily close to capacity for some rate, as well as
beating the $1-\sqrt{R}$ bound for every rate, both remained
open\footnote{Independent of our work, Alex Vardy (personal
  communication) constructed a variant of the code defined
  in~\cite{PV-focs05} which could be list decoded with fraction of
  errors more than $1-\sqrt{R}$ for all rates $R$. However, his
  construction gives only a small improvement over the $1-\sqrt{R}$
  bound and does not achieve the list decoding capacity.}.

\subsection{Our Results}

In this paper, we describe codes that get arbitrarily close to the
list decoding capacity $\p_{\rm cap}(R)$ for every rate.  In other
words, we give explicit codes of rate $R$ together with polynomial
time list decoding up to a fraction $1-R-\eps$ of errors for every
rate $R$ and arbitrary $\eps > 0$. As remarked before, this attains
the information-theoretically best possible trade-off one can hope for
between the rate and error-correction radius. While the focus of our
presentation is primarily on the major asymptotic improvements we
obtain over previous methods, we stress that our results offers a
complexity vs. performance trade-of and gives non-trivial
improvements, even for large rates and modest block lengths, with a
value of the ``folding parameter'' $m$ as small as $4$.
A discussion of the bounds for small values of $m$ appears in
Section~\ref{sec:tri-prac}.

  Our codes are simple to describe: they are {\em folded Reed-Solomon
    codes}, which are in fact {\em exactly} Reed-Solomon (RS) codes, but
    viewed as a code over a larger alphabet by careful bundling of
    codeword symbols. Given the ubiquity of RS codes, this is an
    appealing feature of our result, and in fact our methods directly
    yield better decoding algorithms for RS codes when errors occur in
    {\em phased bursts} (a model considered in \cite{krachkovsky}).


Our result extends easily to the problem of {\em list recovery} (see
Definition~\ref{def:lr}). The biggest advantage here is that we are
able to achieve a rate that is independent of the size of the input
lists.  This is an extremely useful feature in concatenated code
constructions. We are able to use this to reduce the alphabet size
needed to achieve capacity, and also obtain results for binary codes.
We briefly describe these results below.

To get within $\eps$ of capacity, the folded RS codes that we
construct have alphabet size $n^{O(1/\eps)}$ where $n$ is the block
length. By concatenating our codes of rate close to $1$ (that are list
recoverable) with suitable inner codes followed by redistribution of
symbols using an expander graph (similar to a construction for
linear-time unique decodable codes in \cite{GI-ieeejl}), we can get
within $\eps$ of capacity with codes over an alphabet of size
$2^{O(\eps^{-4} \log (1/\eps))}$.  A counting argument shows that
codes that can be list decoded efficiently to within $\eps$ of the
capacity need to have an alphabet size of $2^{\Omega(1/\eps)}$, so the
alphabet size we attain is in the same ballpark as the best possible.

\begin{figure}[h]
\begin{center}
\epsfysize=3in
 \epsffile{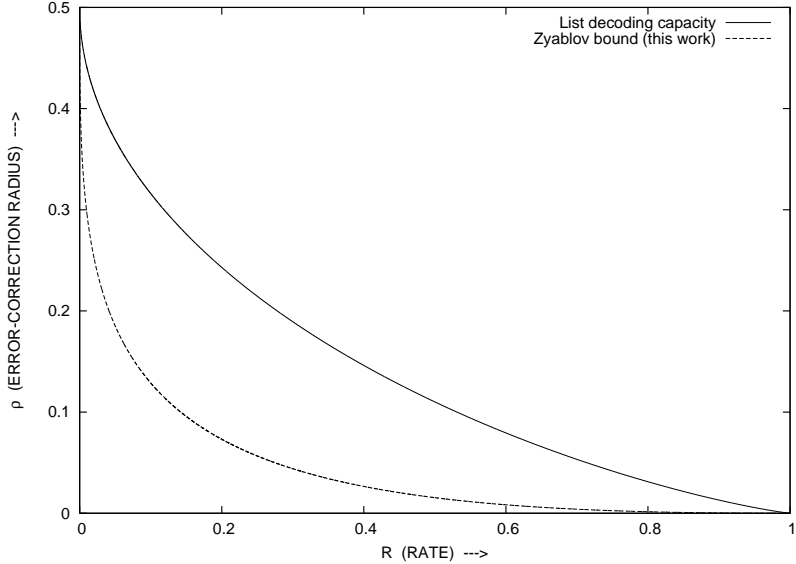}
  \end{center}
  \caption{{\rm Error-correction radius $\p$ of our algorithm for
 binary codes plotted against the rate $R$. The best possible
 trade-off, i.e., capacity, is $\p=H^{-1}(1-R)$, and is also plotted.}}
  \label{fig:binary-ld}
\end{figure}

For binary codes, the list decoding capacity is known to be $\p_{\rm
  bin}(R) = H^{-1}(1-R)$ where $H(\cdot)$ denotes the binary entropy
function~\cite{elias91,GHSZ}. We do not know explicit constructions of
binary codes that approach this capacity. However, using our codes in
a natural concatenation scheme, we give polynomial time constructible
binary codes of rate $R$ that can be list decoded up to a fraction
$\p_{\rm Zyab}(R)$ of errors, where $\p_{\rm Zyab}(R)$ is the
``Zyablov bound''. See Figure~\ref{fig:binary-ld} for a plot of these
bounds.

\subsection{Bibliographic Remarks}

These results were first reported in \cite{GR-stoc06}.  We would like
to point out that the presentation in this paper is somewhat different
from the original papers~\cite{PV-focs05,GR-stoc06} in terms of
technical details, organization, as well as chronology.  With the
benefit of hindsight, we believe this alternate presentation to be
simpler and more self-contained direct than the description in
\cite{GR-stoc06}, which used the results of Parvaresh-Vardy as a
black-box. The exact relationship of our codes to the Parvaresh-Vardy
construction is spelled out in detail in
Section~\ref{sec:pv-relation}. Below, we discuss some technical
aspects of the original development of this material, in order to shed
light on the origins of our work.  We also point the reader to the
survey~\cite{Gur-nowsurvey} for a detailed treatment of recent
advances in algorithms for list decoding.

Two independent works by Coppersmith and
Sudan~\cite{coppersmith-sudan} and Bleichenbacher, Kiayias and
Yung~\cite{BKY07} considered the variant of RS codes where the message
consists of two (or more) independent polynomials over some field
$\F$, and the encoding consists of the joint evaluation of these
polynomials at elements of $\F$ (so this defines a code over
$\F^2$).\footnote{The resulting code is in fact just a Reed-Solomon
  code where the evaluation points belong to the subfield $\F$ of the
  extension field over $\F$ of degree two.} A naive way to decode
these codes, which are also called ``interleaved Reed-Solomon codes,''
would be to recover the two polynomials individually, by running
separate instances of the RS decoder. Of course, this gives no gain
over the performance of RS codes. The hope in these works was that
something can possibly be gained by exploiting that errors in the two
polynomials happen at ``synchronized'' locations.  However, these
works could not give any improvement over the $1-\sqrt{R}$ bound known
for RS codes for worst-case errors.  Nevertheless, for {\em random
  errors}, where each error replaces the correct symbol by a uniform
random field element, they were able to correct well beyond a fraction
$1-\sqrt{R}$ of errors. In fact, as the order of interleaving (i.e.,
number of independent polynomials) grows, the radius approaches the
optimal value $1-R$. Since these are large alphabet codes, this model
of random errors is not interesting from a coding-theoretic
perspective, %
\footnote{This is because, as pointed out by Piotr Indyk, over large
  alphabets one can reduce decoding from uniformly random errors to
  decoding from erasures with a negligible loss in rate. The idea is
  to pad each codeword symbol with a small trail of $0$'s; a uniformly
  random error is highly unlikely to keep each of these $0$'s intact,
  and can thus be detected and declared as an erasure. Now recall that
  decoding from a fraction $1-R$ of erasures with rate $R$ is easy
  using Reed-Solomon codes.}%
though the algorithms are interesting from an algebraic viewpoint.

In \cite{PV-allerton}, Parvaresh and Vardy gave a {\em heuristic} decoding
algorithm for these interleaved RS codes based on multivariate
interpolation. However, the provable performance of these codes
coincided with the $1-\sqrt{R}$ bound for Reed-Solomon codes.  The key
obstacle in improving this bound was the following: for the case when
the messages are pairs $(f(X),g(X))$ of degree $k$ polynomials, two algebraically
independent relations were needed to identify both $f(X)$ and
$g(X)$. The interpolation method could only provide one such relation
in general (of the form $Q(X,f(X),g(X)) = 0$ for a trivariate
polynomial $Q(X,Y,Z)$). This still left too much ambiguity in the
possible values of $(f(X),g(X))$. (The approach in \cite{PV-allerton}
was to find several interpolation polynomials, but there was no
guarantee that they were not all algebraically dependent.)

Then, in \cite{PV-focs05}, Parvaresh and Vardy put forth the ingenious
idea of obtaining the extra algebraic relation essentially ``for
free'' by enforcing it as an {\it a priori} condition satisfied at the
encoder. Specifically, instead of letting the second polynomial $g(X)$
to be an independent degree $k$ polynomial, their insight was to make
it correlated with $f(X)$ by a specific algebraic condition, such as
$g(X) = f(X)^d \mod {E(X)}$ for some integer $d$ and an irreducible
polynomial $E(X)$ of degree $k+1$.


Then, once we have the interpolation polynomial $Q(X,Y,Z)$, $f(X)$ can
be obtained as follows: Reduce the coefficients of
$Q(X,Y,Z)$ modulo $E(X)$ to get a polynomial $T(Y,Z)$ with
coefficients from $\F[X]/(E(X))$ and then find roots of the univariate
polynomial $T(Y,Y^d)$. This was the key idea in \cite{PV-focs05} to
improve the $1-\sqrt{R}$ decoding radius for rates less than
$1/16$. For rates $R \to 0$, their decoding radius approached $1-O(R
\log (1/R))$.

The modification to using independent polynomials, however, does not come for
free. In particular, since one sends at least twice as much
information as in the original RS code, there is no way to construct
codes with rate more than $1/2$ in the PV scheme. If we use $s \ge 2$
correlated polynomials for the encoding, we incur a factor $1/s$ loss
in the rate. This proves quite expensive, and as a result the
improvements over RS codes offered by these codes are only manifest at
very low rates.

The central idea behind our work
is to avoid this rate loss by
making the correlated polynomial $g(X)$ essentially identical to the
first (say $g(X) = f(\gamma X)$). Then the evaluations of $g(X)$ can
be inferred as a simple cyclic shift of the evaluations of $f(X)$, so
intuitively there is no need to explicitly include those too in the
encoding.  

\subsection{Organization}
We begin with a description of our code construction, folded
Reed-Solomon codes, and outline their relation to Parvaresh-Vardy
codes in Section~\ref{sec:foldedRS}. In Section~\ref{sec:trivariate},
we present and analyze a trivariate interpolation based decoder for
folded RS codes, which lets us approach a decoding radius of
$1-R^{2/3}$ with rate $R$. 
In Section~\ref{sec:capacity-codes}, we
extend the approach to $(s+1)$-variate interpolation for any $s \ge
3$, allowing us to decode up to radius $1-R^{s/(s+1)}$, and by picking
$s$ large enough obtain our main result
(Theorem~\ref{thm:final-capacity}) on explicit codes achieving list
decoding capacity. In Section~\ref{sec:extensions}, we generalize our
decoding algorithm to the list recovery setting with almost no loss
in rate, and use this powerful primitive to reduce the alphabet size
of our capacity-achieving codes to a constant depending only on
distance to capacity as well as to construct binary codes
list-decodable up to the Zyablov bound. Finally, we close with some
remarks in Section~\ref{sec:concl}.

\section{Folded Reed-Solomon Codes}
\label{sec:foldedRS}

In this section, we will use a simple variant of Reed-Solomon codes
called folded Reed-Solomon codes for which we can beat the
$1-\sqrt{R}$ decoding radius possible for RS codes. In fact, by
choosing parameters suitably, we can decode close to the optimal
fraction $1-R$ of errors with rate $R$.

\subsection{Description of Folded Codes}
Consider a Reed-Solomon code $C' = {\sf RS}_{\F,\F^*}[n',k]$ consisting
of evaluations of degree $k$ polynomials over $\F$ at the set $\F^*$
of nonzero elements of $\F$. Let $q = |\F| = n'+1$. Let $\g$ be a
generator of the multiplicative group $\F^*$, and let the evaluation
points be ordered as $1,\g,\g^2,\dots,\g^{n'-1}$.  Using all nonzero
field elements as evaluation points is one of the most commonly used
instantiations of Reed-Solomon codes.

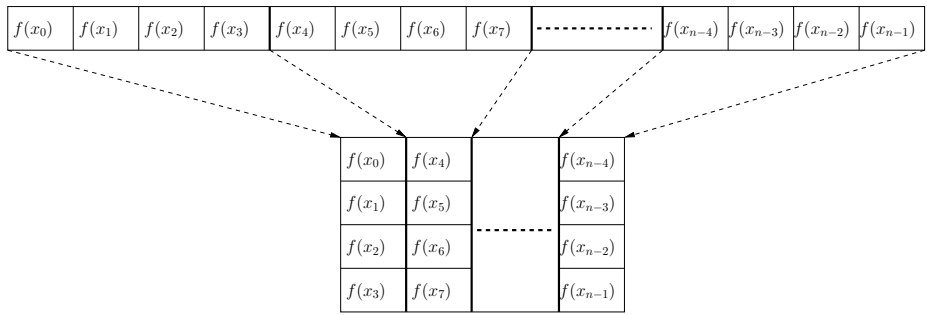
\begin{figure*}[h]
\begin{center}
\resizebox{4.9in}{!}{\input{frs.pstex_t}}
\caption{Folding of the Reed Solomon code with parameter $m=4$.}
\label{fig:frs}
\end{center}
\end{figure*}

Let $m \ge 1$ be an integer parameter called the {\em folding
  parameter}. Define $n\le n'$ to be the largest integer that is divisible by
$m$. Let $C$ be the $[n,k]_{\F}$ RS code that is defined by the set of
evaluation points $1,\g,\g^2,\dots,\g^{n-1}$. In other words, $C$ is obtained
from $C'$ by truncating the last $n'-n$ symbols.
Note that $m$
divides $n$. 
\begin{definition}[Folded Reed-Solomon Code]
  The $m$-folded version of the RS code $C$, denoted
  $\frs_{\F,\gamma,m,k}$, is a code of block length $N = n/m$ over
  $\F^m$, where $n \le |\F|-1$ is the largest integer that is divisible by $m$. The encoding of a message $f(X)$, a
  polynomial over $\F$ of degree at most $k$, has as its $j$'th
  symbol, for $0 \le j < n/m$, the $m$-tuple $(f(\gamma^{jm}),
  f(\gamma^{jm+1}), \cdots , f(\gamma^{jm+m-1}))$.  In other words,
  the codewords of $\frs_{\F,\gamma,m,k}$ are in one-one
  correspondence with those of the RS code $C$ and are obtained by
  bundling together consecutive $m$-tuple of symbols in codewords of
  $C$.
%
\end{definition}
We illustrate the above construction for the choice $m=4$ in
Figure~\ref{fig:frs}. The polynomial $f(X)$ is the message, whose
Reed-Solomon encoding consists of the values of $f$ at
$x_0,x_1,\dots,x_{n-1}$ where $x_i = \gamma^i$. Then, we perform a
folding operation by bundling together tuples of $4$ symbols to give a
codeword of length $n/4$ over the alphabet $\F^4$.

Note that the folding operation does not change the rate $R$ of the
original Reed-Solomon code. The relative distance of the folded RS
code also meets the Singleton bound and is at least $1-R$. 

\begin{remark}[Origins of term ``folded RS codes'']
The terminology of
folded RS codes was coined in \cite{krachkovsky}, where an algorithm
to correct random errors in such codes was presented (for a noise
model similar to the one used in \cite{coppersmith-sudan,BKY07} that was
mentioned earlier). The motivation was to decode RS codes from many
random ``phased burst'' errors. Our decoding algorithm for folded RS
codes can also be likewise viewed as an algorithm to correct beyond
the $1-\sqrt{R}$ bound for RS codes if errors occur in large, phased
bursts (the actual errors can be adversarial).
\end{remark}

\subsection{Why might folding help?}
Since folding seems like such a simplistic operation, and the
resulting code is essentially just a RS code but viewed as a code over
a large alphabet, let us now understand why it can possibly give hope
to correct more errors compared to the bound for RS codes.

Consider the folded RS code with folding parameter $m=4$.  First of
all, decoding the folded RS code up to a fraction $p$ of errors is
certainly not harder than decoding the RS code up to the same fraction
$p$ of errors. Indeed, we can ``unfold'' the received word of the
folded RS code and treat it as a received word of the original RS code
and run the RS list decoding algorithm on it. The resulting list will
certainly include all folded RS codewords within distance $p$ of the
received word, and it may include some extra codewords which we can,
of course, easily prune.

In fact, decoding the folded RS code is a strictly easier task.  It is
not too hard to see that correcting $mT$ errors, where the errors
occur in $T$ contiguous blocks involves far few error patterns than
correcting $mT$ errors that can be arbitrarily distributed.  As a
concrete example, say we want to correct a fraction $1/4$ of
errors. Then, if we use the RS code, our decoding algorithm ought to
be able to correct an error pattern that corrupts every $4$'th symbol
in the RS encoding of $f(X)$ (i.e., corrupts $f(x_{4i})$ for $0 \le i
< n/4$). However, after the folding operation, this error pattern
corrupts every one of the symbols over the larger alphabet $\F^4$, and
thus need not be corrected. In other words, for the same fraction of
errors, the folding operation reduces the total number of error
patterns that need to be corrected, since the channel has less
flexibility in how it may distribute the errors.

It is of course far from clear how one may exploit this to actually
correct more errors. To this end, algebraic ideas that exploit the
specific nature of the folding and the relationship between a
polynomial $f(X)$ and its shifted counterpart $f(\gamma X)$ will be
used. These will become clear once we describe our algorithms later
in the paper.

We note that above ``simplification'' of the channel is not attained
for free since the alphabet size increases after the folding
operation\footnote{However, we note that most of the operations in
  decoding still take place in the original field.}. For folding
parameter $m$ that is an absolute constant, the increase in alphabet
size is moderate and the alphabet remains polynomially large in the
block length. (Recall that the RS code has an alphabet size that is
linear in the block length.)  Still, having an alphabet size that is a
large polynomial is somewhat unsatisfactory. Fortunately, existing
alphabet reduction techniques, which are used in
Section~\ref{sec:const-alphabets}, can handle polynomially large
alphabets, so this does not pose a big problem. Moreover, the benefits
of our results kick in already for very small values of $m$ (see
Section~\ref{sec:tri-prac}).

\subsection{Relation to Parvaresh Vardy codes}
\label{sec:pv-relation}
In this subsection, we relate folded RS codes to the
Parvaresh-Vardy (PV) codes~\cite{PV-focs05}, which among other things
will help make the ideas presented in the previous subsection more concrete. 

The basic idea in the PV
codes is to encode a polynomial $f$ by the evaluations of $s \ge 2$
polynomials $f_0=f,f_1,\dots,f_{s-1}$ where $f_i(X) = f_{i-1}(X)^d
\mod E(X)$ for an appropriate power $d$ (and some irreducible polynomial $E(X)$) --- let us call $s$ the order
of such a code. 
Our first main idea is to pick the irreducible polynomial $E(X)$ (and the
parameter $d$) in such a 
manner that every polynomial $f$ of degree at most $k$ satisfies the following
identity: $f(\gamma X)=f(X)^d\mod E(X)$, where $\gamma$ is the generator
of the underlying field.
Thus, a folded RS code with
bundling using an $\gamma$ as above is in fact exactly the PV code of
order $s=m$ for the set of evaluation points
$\{1,\gamma^m,\gamma^{2m},\dots,\gamma^{(n/m-1)m}\}$. 
This is nice as it shows that
PV codes can meet the Singleton bound (since folded RS codes do), but
as such does not lead to any better codes for list decoding.

\begin{figure}[h]
\begin{center}
\resizebox{6in}{!}{\input{reduction-journal.pstex_t}}
\caption{The correspondence between a folded Reed-Solomon code (with $m=4$ and
$x_i=\gamma^i$)
and the Parvaresh Vardy code (of order $s=2$)
evaluated over $\{1,\gamma,\gamma^2,\gamma^{4},\dots,\gamma^{n-4},\dots,
\gamma^{n-2}\}$. 
The 
correspondence for the first block in the folded RS codeword and the first
three blocks in the {\rm PV} codeword is shown explicitly in the left corner
of the figure.}
\label{fig:redux}
\end{center}
\end{figure}
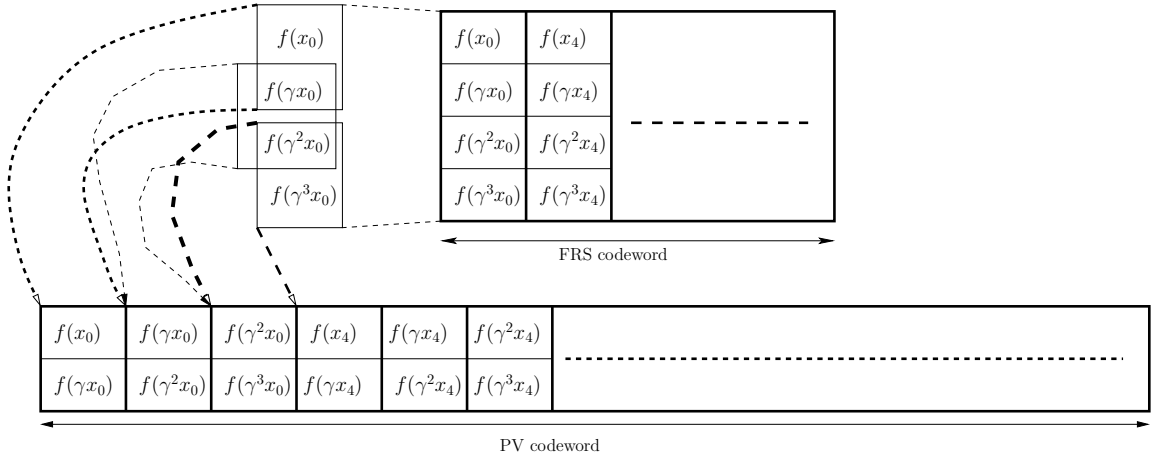

Here comes our second main idea. Let us compare the folded RS code to a PV
code of order $2$ (instead of order $m$) for the set of evaluation
points $\{1,\gamma,\dots\gamma^{m-2},\gamma^{m},\dots,\gamma^{n-m},\dots,
\gamma_{n-2}\}$. We find that in the PV encoding of
$f$, for every $0\le i\le n/m-1$ and every $0< j<m-1$,  $f(\gamma^{mi+j})$ 
appears exactly twice (once as $f(\gamma^{mi+j})$ and another
time as $f_1(\gamma^{-1} \gamma^{mi+j})$), whereas it appears only once
in the folded RS encoding. (See Figure~\ref{fig:redux} for an example
when $m=4$ and $s=2$.)
In other words, the PV and folded RS codes have the
same information, but the rate of the folded RS codes is bigger by a factor
of $\frac{2m-2}{m}=2-\frac{2}{m}$. 
Decoding the folded RS codes from a fraction $\rho$ of errors
reduces to correcting the same fraction $\rho$ of errors for the PV
code. But the rate vs. error-correction radius trade-off is better for
the folded RS code since it has (for large enough $m$, almost) twice  
the rate of the PV code.

In other words, our folded RS codes are chosen such that they are
``compressed'' forms of suitable PV codes, and thus have better rate
than the corresponding PV code for a similar error-correction
performance. This is where our gain is, and using this idea we are
able to construct folded RS codes of rate $R$ that are list decodable
up to radius roughly $1- \sqrt[s+1]{R^s}$ for any $s \ge 1$. Picking
$s$ large enough lets us get within any desired $\eps$ from capacity.

\section{Trivariate interpolation based decoding}
\label{sec:trivariate}
The list decoding algorithm for RS codes from \cite{sudan,GS98} is
based on bivariate interpolation. The key factor driving the agreement
parameter $t$ needed for the decoding to be successful was the
($(1,k)$-weighted) degree $D$ of the interpolated bivariate
polynomial. Our quest for an improved algorithm for folded RS codes
will be based on trying to lower this degree $D$ by using more degrees
of freedom in the interpolation.  Specifically, we will try to use
{\em trivariate interpolation} of a polynomial $Q(X,Y_1,Y_2)$ through
$n$ points in $\F^3$. This enables performing the interpolation with
$D = O(\sqrt[3]{k^2 n})$, which is much smaller than the
$\Theta(\sqrt{kn})$ bound for bivariate interpolation. In principle,
this could lead to an algorithm that works for agreement fraction
$R^{2/3}$ instead of $R^{1/2}$. Of course, this is a somewhat simplistic
hope and
additional ideas are needed to make this approach work. We now
turn to the task of developing a trivariate interpolation based
decoder and proving that it can indeed decode up to a fraction
$1-R^{2/3}$ of errors.

\subsection{Facts about trivariate interpolation}
We begin with some basic definitions and facts
concerning trivariate polynomials.
%
\begin{definition}
\label{def:triv-1}
  For a polynomial $Q(X,Y_1,Y_2) \in \F[X,Y_1,Y_2]$, its
  $(1,k,k)$-weighted degree is defined to be the maximum value of
  $\ell + k j_1 + k j_2$ taken over all monomials $X^\ell Y_1^{j_1}
  Y_2^{j_2}$ that occur with a nonzero coefficient in $Q(X,Y_1,Y_2)$.
\end{definition}
\begin{definition}[Multiplicity of zeroes]
\label{def:triv-2}
  A polynomial $Q(X,Y_1,Y_2)$ over $\F$ is said to have a zero of
  multiplicity $r \ge 1$ at a point $(\a,\b_1,\b_2) \in \F^3$ if
  $Q(X+\a,Y_1+\b_1,Y_2+\b_2)$ has no monomial of degree less than $r$
  with a nonzero coefficient. (The degree of the monomial $X^i
  Y_1^{j_1} Y_2^{j_2}$ equals $i+j_1+j_2$.)
\end{definition}
\begin{lemma}
\label{lem:1-k-k-mult}
Let $\{(\a_i,y_{i1},y_{i2})\}_{i=1}^{n_0}$ be an arbitrary set of $n_0$
triples from $\F^3$.  Let $Q(X,Y_1,Y_2) \in \F[X,Y_1,Y_2]$ be a
nonzero polynomial of $(1,k,k)$-weighted degree at most $D$ that has a
zero of multiplicity $r$ at $(\a_i,y_{i1},y_{i2})$ for every $i$, $1 \le i \le n _0$. Let $f(X),g(X)$ be polynomials of degree at most $k$ such that
for at least $t > D/r$ values of $i$, we have $f(\a_i) =
y_{i1}$ {\em and} $g(\a_i) = y_{i2}$.  Then, $Q(X,f(X),g(X)) \equiv
0$.
\end{lemma}
\begin{proof}
  If we define $R(X) = Q(X,f(X),g(X))$, then $R(X)$ is a univariate
  polynomial of degree at most $D$.  Now, for every $i$ for which
  $f(\a_i) = y_{i1}$ {\em and} $g(\a_i) = y_{i2}$, $(X-\a_i)^r$
  divides $R(X)$ (this follows from the definition of what it means
  for $Q$ to have a zero of multiplicity $r$ at
  $(\a_i,f(\a_i),g(\a_i))$). Therefore if $r t > D$, then $R(X)$ has
  more roots (counting multiplicities) than its degree, and so it must
  be the zero polynomial.
\end{proof}
\begin{lemma}
\label{lem:tri-interpolation}
Given an arbitrary set of $n_0$ triples
$\{(\a_i,y_{i1},y_{i2})\}_{i=1}^{n_0}$ from $\F^3$ and an integer
parameter $r \ge 1$, there exists a nonzero polynomial $Q(X,Y_1,Y_2)$
over $\F$ of $(1,k,k)$-weighted degree at most $D$ such that
$Q(X,Y_1,Y_2)$ has a zero of multiplicity $r$ at
$(\a_i,y_{i1},y_{i2})$ for all $i \in \{1,2\dots,n_0\}$, provided
\begin{equation}
\label{eq:tri-deg-bound}
\frac{D^3}{6k^2} > n_0 {{r+2} \choose 3} \ . 
\end{equation} 
Moreover, we can find such a $Q(X,Y_1,Y_2)$ in time polynomial in
$n_0,r$ by solving a system of homogeneous linear equations over $\F$.
\end{lemma}
\begin{proof}
  The condition that $Q(X,Y_1,Y_2)$ has a zero of multiplicity $r$ at
  a point amounts to ${{r+2} \choose 3}$ homogeneous linear conditions
  in the coefficients of $Q$. The number of monomials in
  $Q(X,Y_1,Y_2)$ equals the number, say $N_3(k,D)$, of triples
  $(i,j_1,j_2)$ of nonnegative integers that obey $i + k j_1 + k j_2
  \le D$. One can show that the number $N_3(k,D)$ is at least as large
  as the volume of the 3-dimensional region $\{ x + k y_1 + k y_2 \le
  D \mid x,y_1,y_2 \ge 0\} \subset \R^3$~\cite{PV-focs05}. An easy
  calculation shows that the latter volume equals $\frac{D^3}{6 k^2}$.
  Hence, if $\frac{D^3}{6k^2} > n_0 {{r+2} \choose 3}$, then the
  number of unknowns exceeds the number of equations, and we are
  guaranteed a nonzero solution. (See Remark~\ref{remark:triv-degree}
  for an accurate estimate of the number of monomials of
  $(1,k,k)$-weighted degree at most $D$, which sometimes leads to a
  better condition under which a polynomial $Q$ with the stated
  property exists.)
\end{proof}

\subsection{Using trivariate interpolation for Folded RS codes}
\label{sec:tri-algo}
Let us now see how trivariate interpolation can be used in the context
of decoding the folded RS code $C'' = \frs_{\F,\gamma,m,k}$ of block
length $N = n/m$. (Throughout this section, we will use
$n$ to denote the block length of the ``unfolded" RS code.) Given a received word $\mv{z} \in (\F^m)^N$ for $C''$
that needs to be list decoded, we define $\mv{y} \in \F^n$ to be the
corresponding ``unfolded'' received word. (Formally, let the $j$'th
symbol of $\mv{z}$ be $(z_{j,0},\dots,z_{j,m-1})$ for $0 \le j < N$.
Then $\mv{y}$ is defined by $y_{jm+l} = z_{j,l}$ for $0 \le j < N$ and
$0 \le l < m$.) Finally define $I$ to be the set $\{0,1,2,\dots,n-1\}\setminus
\{m-1,2m-1,\dots,n-1\}$ and let $n_0=|I|$. Note that $n_0=(m-1)n/m$.

Suppose $f(X)$ is a polynomial whose encoding agrees with $\mv{z}$ on
at least $t$ locations. Then, here is an obvious but important
observation:
\begin{quote} 
For at least $t(m-1)$ values of $i$, $i\in I$, {\em both} the
equalities $f(\g^i) = y_i$ and $f(\gamma^{i+1}) = y_{i+1}$ hold.
\end{quote}
Define the notation $g(X) = f(\gamma X)$. Therefore, if we consider
the $n_0$ triples $(\gamma^i,y_i,y_{i+1}) \in \F^3$ for
$i\in I$, then for at least
$t(m-1)$ triples, we have $f(\g^i) = y_i$ {\em and} $g(\g^i) =
y_{i+1}$. This suggests that interpolating a polynomial $Q(X,Y_1,Y_2)$
through these $n_0$ triples and employing Lemma~\ref{lem:1-k-k-mult}, we
can hope that $f(X)$ will satisfy $Q(X,f(X),f(\g X)) = 0$, and then
somehow use this to find $f(X)$. We formalize this in the following
lemma. The proof follows immediately from the preceding discussion and
Lemma~\ref{lem:1-k-k-mult}.
\begin{lemma}
\label{lem:trivariate-main}
Let $\mv{z} \in (\F^m)^N$ and let $\mv{y} \in \F^n$ be the unfolded
version of $\mv{z}$.  Let $Q(X,Y_1,Y_2)$ be any nonzero polynomial
over $\F$ of $(1,k,k)$-weighted degree at $D$ that has a zero of
multiplicity $r$ at $(\g^i,y_i,y_{i+1})$ for $i\in I$. Let
$t$ be an integer such that $t > \frac{D}{(m-1) r}$.  Then every
polynomial $f(X) \in \F[X]$ of degree at most $k$ whose encoding
according to $\frs_{\F,\g,m,k}$ agrees with $\mv{z}$ on at least $t$
locations satisfies $Q(X,f(X),f(\g X)) \equiv 0$.
\end{lemma}
Lemmas~\ref{lem:tri-interpolation} and \ref{lem:trivariate-main}
motivate the following approach to list decoding the folded RS code
$\frs_{\F,\g,m,k}$. Here $\mv{z} \in (\F^m)^N$ is the received word
and $\mv{y}=(y_0,y_1,\dots,y_{n-1}) \in \F^n$ is its unfolded
version. The algorithm uses an integer multiplicity parameter $r \ge
1$, and is intended to work for an agreement parameter $1 \le t \le
N$.

\medskip
\noindent Algorithm {\sf Trivariate-FRS-decoder}:
\begin{description}
\item[Step 1] (Trivariate Interpolation) Define the degree parameter
\begin{equation}
\label{eq:def-D}
D = \lfloor
  \sqrt[3]{k^2 n_0 r(r+1)(r+2)} \rfloor + 1 \ . 
\end{equation} Interpolate a nonzero
  polynomial $Q(X,Y_1,Y_2)$ with coefficients from $\F$ with the
  following two properties: (i) $Q$ has $(1,k,k)$-weighted degree at
  most $D$, and (ii) $Q$ has a zero of multiplicity $r$ at
  $(\g^i,y_i,y_{i+1})$ for $i\in I$.
  (Lemma~\ref{lem:tri-interpolation} guarantees the feasibility of
  this step as well as its  
  computability in time polynomial in $r$ and $n_0$ (and hence, $n$).)
\item[Step 2] (Trivariate ``Root-finding'') Find a list of all degree $\le k$ polynomials $f(X) \in \F[X]$ such that $Q(X,f(X),f(\gamma X)) = 0$. Output those whose encoding agrees with $\mv{z}$ on at least $t$ locations.
\end{description}

Ignoring the time complexity of Step 2 for now, we can already claim
the following result concerning the error-correction performance of this
strategy.
\begin{theorem}
\label{thm:trivariate-algo-perf}
The algorithm {\sf Trivariate-FRS-decoder} successfully list decodes
the folded Reed-Solomon code $\frs_{\F,\g,m,k}$ up to a number of
errors equal to $\left( N - \left\lfloor N
  \sqrt[3]{\left(\frac{mk}{(m-1)n}\right)^2 \left(1+\frac1r\right)
    \left(1+\frac2r\right)} \right\rfloor - 2\right)$.
\end{theorem}
\begin{proof}
By Lemma~\ref{lem:trivariate-main}, we know that any $f(X)$ whose
encoding agrees with $\mv{z}$ on $t$ or more locations will be output
in Step 2, provided $t > \frac{D}{(m-1) r}$.  For the choice of $D$ in
(\ref{eq:def-D}), this condition is met for the choice $t = 1 +  \lfloor
  \sqrt[3]{\frac{k^2 n_0}{(m-1)^3} \left(1 + \frac{1}{r} \right) \left(
      1 + \frac2r \right)} + \frac{1}{(m-1)r} \rfloor$.  The number of errors 
is equal to $N-t$, and recalling that $n = m N$ and $n_0=(m-1)n/m$, 
we get the claimed bound on the list decoding radius.
\end{proof}

The rate of the folded Reed-Solomon code is $R = (k+1)/n > k/n$, and
so the fraction of errors corrected (for large enough $r$) is $1 -
\left(\frac{mR}{m-1}\right)^{2/3}$. Note that for $m=2$, this is just
the bound $1- (2R)^{2/3}$ that Parvaresh-Vardy obtained for decoding
their codes using trivariate interpolation~\cite{PV-focs05}. The bound
becomes better for larger values of $m$, and letting the folding
parameter $m$ grow, we can approach a decoding radius of $1 - R^{2/3}$.

\subsection{Root-finding step}

In light of the above discussion in Section~\ref{sec:tri-algo}, the
only missing piece in our decoding algorithm is an {\em efficient} way to
solve the following trivariate ``root-finding'' type problem:
\begin{quote}
  Given a nonzero polynomial $Q(X,Y_1,Y_2)$ with coefficients from a
  finite field $\F$ of size $q$, a primitive element $\gamma$ of the
  field $\F$, and an integer parameter $k < q-1$, find a list of all
  polynomials $f(X)$ of degree at most $k$ such that
  $Q(X,f(X),f(\gamma X)) \equiv 0$.
\end{quote}

\noindent The following simple algebraic lemma is at the heart of our solution
to this problem.
\begin{lemma}
\label{lem:key-algebra}
Let $\F$ be the field $\F_q$ of size $q$, and let $\g$ be a primitive element that generates its multiplicative group. Then we have the following two facts:
\begin{enumerate}
\item The polynomial $E(X) \eqdef X^{q-1} - \g$ is irreducible over $\F$.
\item Every polynomial $f(X) \in \F[X]$ of degree less than $q-1$ satisfies $f(\g X) = f(X)^q \mod E(X)$. 
\end{enumerate}
\end{lemma}
\begin{proof}
The fact that $E(X) = X^{q-1} - \gamma$ is irreducible over $\F_q$
follows from a known, precise characterization of all irreducible
binomials, i.e., polynomials of the form $X^a - c$, see for instance
\cite[Chap. 3, Sec. 5]{LN86}. For completeness, and since this is an
easy special case, we now prove this fact. Suppose $E(X)$ is not
irreducible and some irreducible polynomial $f(X) \in \F[X]$ of degree
$b$, $1 \le b < q-1$, divides it. Let $\zeta$ be a root of $f(X)$ in
the extension field $\F_{q^b}$. We then have $\zeta^{q^b-1} =
1$. Also, $f(\zeta) = 0$ implies $\zeta^{q-1} = \gamma$. These
equations together imply $\gamma^{\frac{q^b-1}{q-1}} = 1$.  Now,
$\gamma$ is primitive in $\F_q$, so that $\gamma^m = 1$ iff $m$ is
divisible by $(q-1)$. We conclude that $q-1$ must divide
$1+q+q^2+\cdots+q^{b-1}$. This is, however, impossible since
$1+q+q^2+\cdots+q^{b-1} \equiv b \pmod{(q-1)}$ and $0 < b < q-1$. This
contradiction proves that $E(X)$ has no such factor of degree less
than $q-1$, and is therefore irreducible.

For the second part, we have the simple but useful identity $f(X)^q =
f(X^q)$ that holds for all polynomials in $\F_q[X]$. Therefore,
$f(X)^q - f(\gamma X) = f(X^q) - f(\gamma X)$. The latter polynomial
is clearly divisible by $X^q - \gamma X$, and thus also by $X^{q-1} -
\gamma$. Hence $f(X)^q \equiv f(\gamma X) \pmod{E(X)}$ which implies
that $f(X)^q \mod E(X) = f(\gamma X)$ since the degree of $f(\gamma
X)$ is less than $q-1$.
\end{proof}

Armed with this lemma, we are ready to tackle the trivariate
root-finding problem.
\begin{theorem}
\label{thm:trf-soln}
There is a deterministic algorithm that on input a finite field $\F$
of size $q$, a primitive element $\gamma$ of the field $\F$, a nonzero
polynomial $Q(X,Y_1,Y_2) \in \F[X,Y_1,Y_2]$ of degree less than $q$ in
$Y_1$, and an integer parameter $k < q-1$, outputs a list of all
polynomials $f(X)$ of degree at most $k$ satisfying the condition
$Q(X,f(X),f(\gamma X)) \equiv 0$. The algorithm has runtime polynomial
in $q$.
\end{theorem}
\begin{proof}
Let $E(X) = X^{q-1} - \gamma$. We know by Lemma~\ref{lem:key-algebra} that $E(X)$ is irreducible. We first divide out the largest power of $E(X)$ that divides $Q(X,Y_1,Y_2)$ to obtain $Q_0(X,Y_1,Y_2)$ where $Q(X,Y_1,Y_2) = E(X)^b Q_0(X,Y_1,Y_2)$ for some $b \ge 0$ and $E(X)$ does not divide $Q_0(X,Y_1,Y_2)$. Clearly, if $f(X)$ satisfies $Q(X,f(X),f(\gamma X)) = 0$, then $Q_0(X,f(X),f(\gamma X)) = 0$ as well, so we will work with $Q_0$ instead of $Q$.
  Let us view $Q_0(X,Y_1,Y_2)$ as a polynomial $T_0(Y_1,Y_2)$ with coefficients
  from $\F[X]$. Further, reduce each of the coefficients modulo $E(X)$ to
  get a polynomial $T(Y_1,Y_2)$ with coefficients from the extension field
  $\tilde{\F} \eqdef \F[X]/(E(X))$ (this is a field since $E(X)$ is
  irreducible over $\F$).  We note that $T(Y_1,Y_2)$ is a nonzero polynomial since $Q_0(X,Y_1,Y_2)$ is not divisible by $E(X)$. 

In view of Lemma~\ref{lem:key-algebra}, it suffices to find degree
  $\le k$ polynomials $f(X)$ satisfying $Q_0(X,f(X),f(X)^q) \pmod
  {E(X)} = 0$. In turn, this means it suffices to find elements
  $\Gamma \in \tilde{\F}$ satisfying $T(\Gamma,\Gamma^q) = 0$. If we
  define the univariate polynomial $R(Y_1) \eqdef T(Y_1,Y_1^q)$, this is
  equivalent to finding all $\Gamma \in \tilde{\F}$ such that
  $R(\Gamma) = 0$, or in other words the roots in $\tilde{\F}$ of
  $R(Y_1)$. 

Now $R(Y_1)$ is a nonzero polynomial since $R(Y_1) =0$ iff $Y_2 -
Y_1^q$ divides $T(Y_1,Y_2)$, and this cannot happen as $T(Y_1,Y_2)$
has degree less than less than $q$ in $Y_1$. The degree of $R(Y_1)$ is
at most $dq$ where $d$ is the total degree of $Q(X,Y_1,Y_2)$. The
characteristic of $\tilde{\F}$ is at most $q$, and its degree over the
base field is at most $q \lg q$. Therefore, we can find all roots of
$R(Y_1)$ by a deterministic algorithm running in time polynomial in
$d,q$~\cite{berl-factor}. Each of the roots will be a polynomial in
$\F[X]$ of degree less than $q-1$. Once we find all the roots, we
prune the list and only output those roots of $f(X)$ that have degree at
most $k$ and satisfy $Q_0(X,f(X),f(\gamma X)) = 0$.
\end{proof}

With this, we have a polynomial time implementation of the algorithm
{\sf Trivariate-FRS-decoder}. There is the technicality that the
degree of $Q(X,Y_1,Y_2)$ in $Y_1$ should be less than $q$. This degree
is at most $D/k$, which by the choice of $D$ in (\ref{eq:def-D}) is at
most $(r+3)\sqrt[3]{n/k} < (r+3) q^{1/3}$. For a fixed $r$ and growing
$q$, the degree is much smaller than $q$. (In fact, for constant rate
codes, the degree is a constant independent of $n$.) By letting $m,r$
grow in Theorem~\ref{thm:trivariate-algo-perf}, and recalling that the
running time is polynomial in $n,r$, we can conclude the following
main result of this section.
\begin{theorem}
\label{thm:final-trivariate}
For every $\epsilon > 0$ and $R$, $0 < R < 1$, there is a family of
$m$-folded Reed-Solomon codes for $m = O(1/\eps)$ that have rate at
least $R$ and which can be list decoded up to a fraction $1 -
(1+\eps)R^{2/3}$ of errors in time polynomial in the block length and
$1/\eps$.
\end{theorem}

\subsection{Alternate decoding bound for high rates and practical considerations}
\label{sec:tri-prac}

In the discussion above, the fraction of errors
$1-\left(\frac{mR}{m-1}\right)^{2/3}$, call it
$\p_{\high}^{(m,2)}(R)$, approaches $1-R^{2/3}$ (and hence improves
upon the bound of $\p_{\GS}(R)=1-\sqrt{R}$ in~\cite{GS98}) for every
rate $R$ for \textit{large enough} $m$.  For practical implementations
the parameter $m$ will be some small fixed integer. Note that for
fixed $m$, the bound of $\p_{\high}^{(m,2)}(R)$ is useless for $R\ge
1-\frac{1}{m}$, whereas the $1-\sqrt{R}$ bound for decoding
Reed-Solomon codes~\cite{GS98} is meaningful for all $R< 1$.  

Given that one is often interested in high rate codes, this suggests
that in order to reap the benefits of our new codes for large rates,
the folding parameter needs to be picked large enough. Fortunately,
this is not the case, and we now show that one can beat the
$1-\sqrt{R}$ bound for all rates $R$ for a {\em fixed} value of the
folding parameter $m$; in fact, a value as small as $m=5$ suffices.
These bounds also hint at the fact that the improvements offered by
the decoding algorithms in this paper are not just asymptotic and kick
in for parameter choices that could be practical.


Our goal now is to sketch how a minor change to the algorithm in
Section~\ref{sec:tri-algo} allows us to correct a fraction
\begin{equation}
\label{eq:high-rate}
\p_{\low}^{(m,2)}(R)=\frac{m}{m+1}\left(1-R^{2/3}\right) 
\end{equation}
of errors. The bound of $\p_{\low}^{(m,2)}(R)$ gives a larger decoding
radius than $\p_{\high}^{(m,2)}(R)$ for large rates $R$. A more
precise comparison of the bounds
$\p_{\high}^{(m,2)},\p_{\low}^{(m,2)}$ and $\p_{\GS}$ is done at the
end of this subsection. The improvement of the decoding radius to
$\p_{\low}^{(m,2)}(R)$ for large rates (and hence small error
fractions) comes via another way to analyze (a variant of) the
algorithm in Section~\ref{sec:tri-algo}, which was suggested to us by
J\o rn Justesen. The algorithm is the same as in
Section~\ref{sec:tri-algo} except that the set of interpolating points
is slightly different. In particular in the trivariate interpolating
step, we choose $I=\{0,1,\dots,n-2\}$. Let $n_0=|I|=n-1$.  The crucial
observation here is that an erroneous symbol $z_j\in\F^m$ (for some
position $0 \le j < N$ in the received word $\mv{z}$) translates to at
most $m+1$ errors among the interpolation tuples in the trivariate
interpolation step.  More precisely, given that $0\le e\le N$ is the
number of errors,
\begin{quote} 
For at least $t'=n_0-e(m+1)$ values of $i$, $i\in I$, {\em both} the
equalities $f(\g^i) = y_i$ and $f(\gamma^{i+1}) = y_{i+1}$ hold.
\end{quote}

\noindent By Lemmas~\ref{lem:1-k-k-mult}, \ref{lem:tri-interpolation}
and the degree bound (\ref{eq:def-D}), the algorithm outlined above
will work as long as
\[n_0-e(m+1)> \sqrt[3]{k^2n_0\left(1+\frac{1}{r}\right)\left(1+\frac{2}{r}\right)}+\frac{1}{r}.\]
Recalling that $n_0=n-1<n$, the above is satisfied if
\[n-1-e(m+1)> \sqrt[3]{k^2n\left(1+\frac{1}{r}\right)\left(1+\frac{2}{r}\right)}+\frac{1}{r}.\]
Recalling that $n=Nm$, the above is satisfied if
\[e<\left(\frac{m}{m+1}\right)N\left(1-\sqrt[3]{\left(\frac{k}{n}\right)^2\left(1+\frac{1}{r}\right)\left(1+\frac{2}{r}\right)}\right)-\frac{2}{m+1}.\]
Noting that $m\ge 1$, leads to the following analog of
Theorem~\ref{thm:trivariate-algo-perf}:
\begin{theorem}
\label{thm:trivariate-algo-perf-justesen}
The version of algorithm {\sf Trivariate-FRS-decoder} discussed above,
successfully list decodes the folded Reed-Solomon code
$\frs_{\F,\g,m,k}$ as long as the number of errors is less than
$\left\lfloor\left(\frac{m}{m+1}\right)N\left(1-\sqrt[3]{\left(\frac{k}{n}\right)^2\left(1+\frac{1}{r}\right)\left(1+\frac{2}{r}\right)}\right)\right\rfloor -1$.
\end{theorem}
For large enough $r$, the above implies that rate $R$ folded RS codes
can be list decoded up to a fraction
$\p_{\low}^{(m,2)}(R)=\left(\frac{m}{m+1}\right)\left(1-R^{2/3}\right)$
of errors.

\subsubsection*{Comparison of the bounds}
\begin{figure}[h]
\begin{center}
\epsfysize=3in
 \epsffile{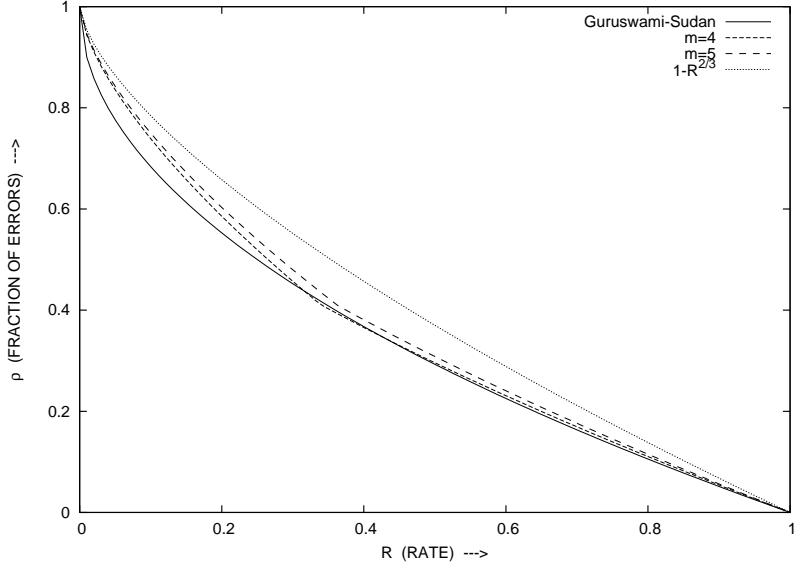}
  \end{center}
  \caption{{\rm Error-correction radius $\max(\p_{\low}^{(m,2)}(R),\p_{\high}^{(m,2)}(R))$ for $m=4,5$. For comparison $\p_{\GS}(R)=1-\sqrt{R}$ and the limit $1-R^{2/3}$ are also plotted. For $m=5$,
the performance of the trivariate interpolation algorithm strictly improves upon
that of $\p_{\GS}$ for all rates.}}
  \label{fig:compare}
\end{figure}

We now make a comparison between the bounds $\p_{\low}^{(m,2)}$,
$\p_{\high}^{(m,2)}$ and $\p_{\GS}$. We first note that
$\p_{\low}^{(m,2)}(R)\ge\p_{\high}^{(m,2)}(R)$ for every rate $R\ge
\left(1-\frac{1}{m}\right)\left(\frac{1}{m+1-\sqrt[3]{m(m-1)^2}}\right)^{3/2}$. In
particular, $\p_{\low}^{(m,2)}(R)>\p_{\high}^{(m,2)}(R)$ for all rates
$R\ge 1-\frac{1}{m}$. Let us now compare $\p_{\low}^{(m,2)}(R)$ and
$\p_{\GS}(R)$.  Specifically, we give a heuristic argument to show
that for high enough rates $R$ and $m\ge 4$, $\p_{\low}^{(m,2)}(R)>
\p_{\GS}(R)$. Let $R=1-\eps$. Then ignoring the $O(\eps^2)$ terms in
the Taylor expansions we get $\p_{\GS}(1-\eps)\approx \eps/2$ and
$\p_{\low}^{(m,2)}(1-\eps)\approx\frac{2m\eps}{3(m+1)}$: the latter
quantity is strictly larger than the former for every $m\ge 4$.
In fact, it can be verified that for all rates $R\ge 0.44$, $\p_{\low}^{(4,2)}>
\p_{\GS}$.
Figure~\ref{fig:compare} plots the tradeoff $\p_{\GS}(R)$ and
$\max\left(\p_{\low}^{(m,2)}(R),\p_{\high}^{(m,2)}(R)\right)$ for some
small values of $m\ge 2$. The limit for large $m$, which is
$1-R^{2/3}$, is also plotted.

\begin{remark}[Better bound on $(1,k,k)$-weighted degree]
\label{remark:triv-degree}
  For small values of the parameter $r$, one should use a better
  estimate for the degree bound $D$ than the bound
  (\ref{eq:tri-deg-bound}) based on the volume argument. The number of monomials $X^i Y_1^{j_1} Y_2^{j_2}$ whose $(1,k,k)$-weighted degree is at most $D$ is exactly equal to 
\begin{equation}
\label{eq:triv-better-bound}
k {{a+2} \choose 3} + (D - a k + 1) {{a+2} \choose 2}
\end{equation}
where $a = \left\lfloor \frac{D}{k} \right\rfloor$. This is often
larger than the $\frac{D^3}{6 k^2}$ lower bound we used in
Lemma~\ref{lem:tri-interpolation}, and certainly for any specific
setting of parameters $n,k,r$, the estimate (\ref{eq:triv-better-bound})
should be used. A similar remark applies for the bound used in
Lemma~\ref{lem:simple-ext} for $(s+1)$-variate interpolation. Since it
makes no difference for the asymptotics, we chose to stick with the
simpler expressions.
%
%
\end{remark}

\section{Codes approaching list decoding capacity}
\label{sec:capacity-codes}
Given that trivariate interpolation improved the decoding radius
achievable with rate $R$ from $1-R^{1/2}$ to $1- R^{2/3}$, it is
natural to attempt to use higher order interpolation to improve the
decoding radius further. In this section, we discuss the (quite
straightforward) technical changes needed for such a generalization.

Consider again the $m$-folded RS code $C' = \frs_{\F,\gamma,m,k}$
where $\F = \F_q$. Let $s$ be an integer in the range $1 \le s \le
m$. We will develop a decoding algorithm based on interpolating an
$(s+1)$-variate polynomial $Q(X,Y_1,Y_2,\dots,Y_s)$. The definitions
of the $(1,k,k,\dots,k)$-weighted degree (with $k$ repeated $s$ times)
of $Q$ and the multiplicity at a point $(\a,\b_1,\b_2,\dots,\b_s) \in
\F^{s+1}$ are straightforward extensions of Definitions
\ref{def:triv-1} and \ref{def:triv-2}.

As before let $\mv{y} = (y_0,y_1,\dots,y_{n-1})$ be the unfolded
version of the received word $\mv{z} \in (\F^m)^N$ of the folded RS
code that needs to be decoded. Define the set of interpolations points to be 
\[I=\{0,1,2,\dots,n-1\}
\setminus\left(\bigcup_{j=0}^{n/m-1}\{jm+m-s+1,jm+m-s+2,\dots,jm+m-1\}\right).\]
The reason for this choice of $I$ is that if the $m$-tuple containing
$y_i$ is correct and $i \in I$, then all the $s$ values
$y_i,y_{i+1},\dots,y_{i+s-1}$ are correct. 

Define $n_0=|I|$. Note that $n_0=n(m-s+1)/m$.  Following algorithm
{\sf Trivariate-FRS-decoder}, for suitable integer parameters $D,r$,
the interpolation phase of the $(s+1)$-variate FRS decoder will fit a
nonzero polynomial $Q(X,Y_1,\dots,Y_s)$ with the following properties:
\begin{enumerate}
\item It has $(1,k,k,\dots,k)$-weighted degree at most $D$
\item It has a zero of multiplicity $r$ at
$(\g^i,y_i,y_{i+1},\dots,y_{i+s-1})$ for $i\in I$.
\end{enumerate}
The following is a straightforward generalization of Lemmas
\ref{lem:tri-interpolation} and \ref{lem:trivariate-main}.
\begin{lemma}
\label{lem:simple-ext}
\begin{enumerate}
\item
Provided $\frac{D^{s+1}}{(s+1)! k^s} >  n_0 {{r+s} \choose {s+1}}$, a nonzero polynomial $Q(X,Y_1,\dots,Y_s)$ with the above stated properties exists and moreover can be found in time polynomial in $n$ and $r^s$.
\item Let $t$ be an integer such that $t > \frac{D}{(m-s+1) r}$.  Then
every polynomial $f(X) \in \F[X]$ of degree at most $k$ whose encoding
according to $\frs_{\F,\g,m,k}$ agrees with the received word $\mv{z}$
on at least $t$ locations satisfies $Q(X,f(X),f(\g X),\dots,f(\g^{s-1} X)) \equiv 0$.
\end{enumerate}
\end{lemma}

\begin{proof}
The first part follows from (i) a simple lower bound on the number of
monomials $X^a Y_1^{b_1} \cdots Y_s^{b_s}$ with $a + k (b_1+b_2+\cdots
+ b_s) \le D$, which gives the number of coefficients of
$Q(X,Y_1,\dots,Y_s)$, and (ii) an estimation of the number of
$(s+1)$-variate monomials of total degree less than $r$, which gives
the number of interpolation conditions per $(s+1)$-tuple.

The second part is similar to the proof of
Lemma~\ref{lem:trivariate-main}. If $f(X)$ has agreement on at least
$t$ locations of $\mv{z}$, then for at least $t(m-s+1)$ of the
$(s+1)$-tuples $(\g^i,y_i,y_{i+1},\dots,y_{i+s-1})$, we have
$f(\g^{i+j}) = y_{i+j}$ for $j=0,1,\dots,s-1$. As in
Lemma~\ref{lem:1-k-k-mult}, we conclude that $R(X) \eqdef
Q(X,f(X),f(\g X),\dots,f(\g^{s-1} X))$ has a zero of multiplicity $r$
at $\g^i$ for each such $(s+1)$-tuple. Also, by design $R(X)$ has
degree at most $D$. Hence if $t(m-s+1)r > D$, then $R(X)$ has more
zeroes (counting multiplicities) than its degree, and thus $R(X) \equiv
0$.
\end{proof}

\bigskip
Note the lower bound condition on $D$ above is met with the choice
\begin{equation}
\label{eq:D-s-var}
D = \left\lfloor \left( k^s n_0 r(r+1) \cdots (r+s) \right)^{1/(s+1)} \right\rfloor + 1  \ .
\end{equation}

The task of finding a list of all degree $k$ polynomials $f(X) \in
\F[X]$ satisfying \\
$Q(X,f(X),f(\gamma X), \dots, f(\gamma^{s-1} X)) =0$ 
can be solved using ideas similar to the proof of
Theorem~\ref{thm:trf-soln}. First, by dividing out by $E(X)$ enough
times, we can assume that not all coefficients of
$Q(X,Y_1,\dots,Y_s)$, viewed as a polynomial in $Y_1,\dots,Y_s$ with
coefficients in $\F[X]$, are divisible by $E(X)$. We can then go
modulo $E(X)$ to get a nonzero polynomial $T(Y_1,Y_2,\dots,Y_s)$ over
the extension field $\tilde{\F} = \F[X]/(E(X))$. Now, by
Lemma~\ref{lem:key-algebra}, we have $f(\gamma^j X) = f(X)^{q^j} \mod
E(X)$ for every $j \ge 1$. Therefore, the task at hand reduces to the
problem of finding all roots $\Gamma \in \tilde{\F}$ of the polynomial
$R(Y_1)$ where $R(Y_1) = T(Y_1,Y_1^q,\dots,Y_1^{q^{s-1}})$. There is
the risk that $R(Y_1)$ is the zero polynomial, but it is easily seen
that this cannot happen if the total degree of $T$ is less than $q$.
This will be the case since the total degree is at most $D/k$, which
is at most $(r+s) (n/k)^{1/(s+1)} \ll q$.

The degree of the polynomial $R(Y_1)$ is at most $q^s$, and therefore
all its roots in $\tilde{\F}$ can be found in $q^{O(s)}$ time. We
conclude that the ``root-finding'' step can be accomplished in
polynomial time.

The algorithm works for agreement $t > \frac{D}{(m-s+1) r}$, which for the choice of $D$ in (\ref{eq:D-s-var}) is satisfied if 
\[ t \ge \frac{(k^s n_0)^{1/(s+1)}}{m-s+1}\left(\prod_{j=1}^s\left(1+\frac{j}{r}\right)\right)^{1/(s+1)} + 2  \ . \]
The above along with the fact that $n_0=N(m-s+1)$ implies
 the following, which is multivariate generalization
of Theorem~\ref{thm:trivariate-algo-perf}.
\begin{theorem}
\label{thm:s-variate-algo-perf}
For every integer $m \ge 1$ and every $s$, $1 \le s \le m$, the
$(s+1)$-variate FRS decoder successfully list decodes the $m$-folded
Reed-Solomon code $\frs_{\F,\g,m,k}$ up to a radius $N-t$ as long as the agreement parameter $t$ satisfies
\begin{equation}
\label{eq:agr-lb-listdecoding}
t \ge \sqrt[s+1]{\left(N \frac{k}{m-s+1}\right)^s \prod_{j=1}^s\left(1+\frac{j}{r}\right)} + 2  \ .
\end{equation}
The algorithm runs in $n^{O(s)}$ time and outputs a list of size at
most $|F|^s=n^{O(s)}$.
\end{theorem}

Recalling that the block length of $\frs_{\F,\g,m,k}$ is $N = n/m$ and the rate
is $(k+1)/n$, the above algorithm can decode a fraction of errors
approaching 
\begin{equation}
\label{eq:m-s-bound}
1 - \sqrt[s+1]{\left(\frac{mR}{m-s+1}\right)^s\prod_{j=1}^s\left(1+\frac{j}{r}\right) } \ 
\end{equation} 
using lists of size at most $q^s$.  By picking $r,m$ large enough
compared to $s$, the decoding radius can be made larger than $1 -
(1+\delta) R^{s/(s+1)}$ for any desired $\delta > 0$. We state this result
formally below.
\begin{theorem}
\label{thm:final-multivariate}
For every $0<\delta \le 1$, integer $s \ge 1$ and $0 < R < 1$, there is
a family of $m$-folded Reed-Solomon codes for $m = O(s/\delta)$ that 
have rate at least $R$ and which can be list decoded up to a fraction
$1 - (1+\delta)R^{s/(s+1)}$ of errors in time $(Nm)^{O(s)}$
 and outputs a list of size at most
$(Nm)^{O(s)}$ where $N$ is the block length of the code. The
alphabet size of the code as a function of the block length $N$ is
$(Nm)^{O(m)}$.
\end{theorem}
\begin{proof}
We first note that (\ref{eq:m-s-bound}) is at least
\begin{equation}
\label{eq:m-s-approx-bound}
1-\left(1+\frac{s}{r}\right)\left(\frac{m}{m-s+1}\right)R^{s/(s+1)}.
\end{equation}
We now instantiate the parameters $r$ and $m$ in terms of $s$ and
$\delta$:
\[r=\frac{3s}{\delta} \qquad m=\frac{(s-1)(3+\delta)}{\delta} \ . \]
With the above choice, we have
\begin{equation*}
\left ( 1+\frac{s}{r}\right)\frac{m}{m-s+1} = \left(1+\frac{\delta}{3}\right)^2 < 1+\delta \ .
\end{equation*}
Together with the bound (\ref{eq:m-s-approx-bound}) on the decoding radius, we conclude that the $(s+1)$-variate decoding algorithm certainly list decodes up to a fraction $1-(1+\delta)R^{s/(s+1)}$ of errors.

The worst case list size is $q^s$ and the claim on the list size follows
by recalling that $q\le n+m$ and $N=n/m$. The alphabet size is $q^m=(Nm)^{O(m)}$.
The running time has two major components: (1) Interpolating the
$s+1$-variate polynomial $Q(\cdot)$, which by Lemma~\ref{lem:simple-ext}
is $(nr^s)^{O(1)}$; and (2) Finding all the roots of the interpolated
polynomial, which takes $q^{O(s)}$ time. Of the two, the time complexity
of the root finding step dominates, which is $(Nm)^{O(s)}$.
\end{proof}

In the limit of large $s$, the decoding radius approaches the list decoding capacity $1-R$, leading to our main result.
\begin{theorem}[Explicit capacity-approaching codes]
\label{thm:final-capacity}
For every $\epsilon > 0$ and $0 < R < 1$, there is a family of folded
Reed-Solomon codes that have rate at least $R$ and which can be list
decoded up to a fraction $1 - R - \eps$ of errors in time (and outputs
a list of size at most)
$(N/\eps^2)^{O(\eps^{-1}\log(1/R))}$ where $N$ is the block length of the code. The alphabet
size of the code as a function of the block length $N$ is
$(N/\eps^2)^{O(1/\eps^2)}$.
\end{theorem}
\begin{proof}
  Given $\eps,R$, we will apply Theorem~\ref{thm:final-multivariate}
  with the choice
\begin{equation}
\label{eq:s-delta-vals}
s=\left\lceil \frac{\log(1/R)}{\log(1+\eps)}\right\rceil \quad \mbox{and} \quad \delta=\frac{\epsilon(1-R)}{R(1+\epsilon)} \  .
\end{equation}
The list decoding radius guaranteed by Theorem~\ref{thm:final-multivariate} is at least
\begin{eqnarray*}
1-(1+\delta)R^{s/(s+1)} & = & 1 - R (1+\delta) (1/R)^{1/(s+1)} \\
& \ge & 1 - R(1+\delta) (1+\eps) ~~\mbox{(by the choice of $s$ in (\ref{eq:s-delta-vals}))} \\
& = & 1 - (R+\eps) ~~\mbox{(using the value of $\delta$)} \ .
\end{eqnarray*}

We now turn our attention to the time complexity of the decoding algorithm
and the alphabet size of the code. To this end we first claim that
$m=O(1/\eps^2)$. To see this note that by the definition of $s$ and
$\delta$: 
\[m= O\left(\frac{s}{\delta}\right) = 
O\left(s\cdot\frac{R(1+\eps)}{\eps(1-R)}\right) =
O\left(\frac{1}{\eps^2}\cdot \frac{R\ln (1/R)}{1-R}\right) =
O(1/\eps^2) \ , \]
where for the last step we used $\ln(1/R)\le \frac{1}{R}-1$ for $0 < R
\le 1$.  The claims on the running time, worst case list size and the
alphabet size of the code follow from
Theorem~\ref{thm:final-multivariate} and the facts that
$m=O(1/\eps^2)$ and $s=O(\eps^{-1}\log(1/R))$.
\end{proof}

\smallskip With the proof of our main theoretical result
(Theorem~\ref{thm:final-capacity}) completed, we close this section
with a few remarks.

\begin{remark}[Improvement to decoding radius for high rates]
  As in Section~\ref{sec:tri-prac}, it is possible to improve the
  bound of (\ref{eq:m-s-bound}) to
\[ \max\left(\frac{m}{m+s-1}\left(1-\sqrt[s+1]{R^s\prod_{j=1}^s\left(1+\frac{j}{s}\right)}\right),
1 - \sqrt[s+1]{\left(\frac{mR}{m-s+1}\right)^s\prod_{j=1}^s\left(1+\frac{j}{r}\right) } \right) \ . \]
The former bound is better for large rates.
\end{remark}

\begin{remark}[Optimality of degree $q$ of relation between $f(X)$ and
  $f(\gamma X)$]
  Let $K$ be the extension field $\F_q[X]/(E(X))$ where $E(X) =
  X^{q-1} - \gamma$. The elements of $K$ are in one-one correspondence
  with polynomials of degree less than $q-1$ over $\F_q$. The content
  of Lemma~\ref{lem:key-algebra}, which we made crucial use of above,
  is that the map $\Gamma : K \rightarrow K$ defined by $f(X) \mapsto
  f(\gamma X)$ is a degree $q$ map over $K$, i.e., as a polynomial
  over $K$, $\Gamma(Z) = Z^q$. The fact that this degree is as large
  as $q$ is in turn the cause for the large list size that we need for
  list decoding. It is natural to ask if a different map $\Gamma'$
  could have lower degree (perhaps over a different extension field
  $K_1$). Unfortunately, it turns out this is not possible, as argued
  below.

  Indeed, let $\Gamma'$ be a ring homomorphism of $\F_q[X]$ defined by
  $\Gamma'(f(X)) = f(G(X))$ for some polynomial $G$ over $\F_q$. Let
  $E_1(X)$ be an irreducible polynomial over $\F_q$ of degree $\ell$,
  and let $K_1 = \F_q[X]/(E_1(X))$ be the associated extension field.
  We can view $\Gamma'$ as a map $\Gamma_1$ on $K_1$ by identifying
  polynomials of degree less than $\ell$ with $K_1$ and defining
  $\Gamma_1(f(X)) = f(G(X)) \mod E_1(X)$.  The key point is that
  $\Gamma_1$ is an {\em $\F_q$-linear} map on $K_1$.  Expressed as a
  polynomial over $K_1$, $\Gamma_1$ must therefore be a {\em
    linearized polynomial}, \cite[Chap. 3, Sec.  4]{LN86}, which has
  only terms with exponents that are powers of $q$ (including $q^0 =
  1$). It turns out that for our purposes $\Gamma_1$ cannot have degree
  $1$, and so it must have degree at least $q$.
\end{remark}

\section{Extensions and Codes over Smaller Alphabets}
\label{sec:extensions}
\subsection{Extension to list recovery}
We now present a very useful generalization of the list decoding
result of Theorem~\ref{thm:final-capacity} to the setting of {\em list
  recovery}. Under the list recovery problem, one is given as
input for each codeword position, not just one but a set of several,
say $l$, alphabet symbols. The goal is to find and output all
codewords which agree with some element of the input sets for several
positions. Codes for which this more general problem can be solved
turn out to be extremely valuable as outer codes in concatenated code
constructions. In short, this
is because one can pass a set of possibilities from decodings of the
inner codes and then list recover the outer code with those sets as
the input. If we only had a list-decodable code at the outer level, we
will be forced to make a unique choice in decoding the inner codes
thus losing valuable information.

\begin{definition}[List Recovery]
\label{def:lr}
A code $C \subseteq \Sigma^n$ is said to be
$(\zeta,l,L)$-list recoverable if for every sequence of sets
$S_1,\dots,S_n$ where each $S_i \subseteq \Sigma$ has at most $l$
elements, the number of codewords $c \in C$ for which $c_i \in S_i$
for at least $\zeta n$ positions $i \in \{1,2,\dots,n\}$ is at most
$L$. 

A code $C \subseteq \Sigma^n$ is said to $(\zeta,l)$-list recoverable
in polynomial time if it is $(\zeta,l,L(n))$-list recoverable for some
polynomially bounded function $L(\cdot)$, and moreover there is a
polynomial time algorithm to find the at most $L(n)$ codewords that
are solutions to any $(\zeta,l,L(n))$-list recovery instance.
\end{definition}
We remark that when $l=1$, $(\zeta,1,\cdot)$-list recovery is the
same as list decoding up to a $(1-\zeta)$ fraction of errors.
List recovery has been implicitly studied in several works; the name
itself was coined in \cite{GI-focs01}.

Theorem~\ref{thm:final-capacity} can be generalized to list recover the
folded RS codes. Specifically, for a FRS code with parameters as in
Section~\ref{sec:capacity-codes}, for an arbitrary constant $l \ge 1$, we can 
 $(\zeta,l)$-list recover
in polynomial time provided
\begin{equation}
\label{eq:list-recover-cond}
\zeta N \ge 
\sqrt[s+1]{\left(\frac{k}{m-s+1}\right)^s\frac{nl}{m}\prod_{j=1}^s\left(1+\frac{j}{r}\right)} + 2  \ .
\end{equation}
where $N=n/m$.  We briefly justify this claim. The generalization of
the list decoding algorithm of Section~\ref{sec:capacity-codes} is
straightforward: instead of one interpolation condition for each
symbol of the received word, we just impose $|S_i| \le l$ many
interpolation conditions for each position $i \in \{1,2,\dots,n\}$
(where $S_i$ is the $i$'th input set in the list recovery instance).
The number of interpolation conditions is at most $n l$, and so
replacing $n$ by $n l$ in the bound of Lemma~\ref{lem:simple-ext}
guarantees successful decoding.  This in turn implies that the
condition on the number of agreement of (\ref{eq:agr-lb-listdecoding})
generalizes to the one in (\ref{eq:list-recover-cond}).  This
straightforward generalization to list recovery is a positive feature
of all interpolation based decoding
algorithms~\cite{sudan,GS98,PV-focs05} beginning with the one due to
Sudan~\cite{sudan}.

\medskip Picking $r \gg s$ and $m \gg s$ in
(\ref{eq:list-recover-cond}), we get $(\zeta,l)$-list recover with
rate $R$ for $\zeta \ge \bigl(l R^s\bigr)^{1/(s+1)}$. Now comes the
remarkable fact: we can pick a suitable $s \gg l$ and perform
$(\zeta,l)$-list recovery with agreement parameter $\zeta \ge R +
\eps$ which is independent of $l$! We state the formal result below
(Theorem~\ref{thm:final-capacity} is a special case when $l = 1$).

\begin{theorem}
\label{thm:main-LR}
For every integer $l\ge 1$, for all $R$, $0 < R < 1$ and $\eps > 0$,
  and for every prime $p$,
  there is an \emph{explicit} family of folded Reed-Solomon codes over
  fields of characteristic $p$ that have rate at least $R$ and which
  can be $(R+\eps,l)$-list recovered in polynomial time.
The alphabet size of a code of block length $N$ in
  the family is $(N/\eps^2)^{O(\eps^{-2} \log{l}/(1-R) )}$.
\end{theorem}
\begin{proof} \textit{(Sketch)} Using the exact same arguments as in the
proof of Theorem~\ref{thm:final-multivariate} to the agreement condition
of (\ref{eq:list-recover-cond}), we get that one can list recover
in polynomial time as long as $\zeta\ge (1+\delta)(lR^s)^{1/(s+1)}$, for
any $\delta>0$. The arguments to obtains a lower bound of $R+\epsilon$
are similar to the ones employed in the proof of 
theorem~\ref{thm:final-capacity}. However, $s$ needs to be defined in a
slightly different manner:
\[s=\left\lceil \frac{\log(l/R)}{\log(1+\eps)}\right\rceil.\]
Also this implies that $m=O\left(\frac{\log l}{(1-R)\eps^2}\right)$, which implies the claimed
bound on the alphabet size of the code.
\end{proof}

\begin{remark}[Soft Decoding]
  The decoding algorithm for folded RS codes from Theorem
  \ref{thm:final-capacity} can be further generalized to handle soft
  information, where for each codeword position $i$ the decoder is
  given as input a non-negative weight $w_{i,z}$ for each possible
  alphabet symbol $z$. The weights $w_{i,z}$ can be used to encode the
  confidence information concerning the likelihood of the the $i$'th
  symbol of the codeword being $z$~\cite{KV}. For any $\eps > 0$,
  for suitable choice of parameters, our codes of rate $R$ over
  alphabet $\Sigma$ have a soft decoding algorithm that outputs all
  codewords $c = \langle c_1,c_2,\dots,c_N \rangle$ that satisfy
\[ \sum_{i=1}^N w_{i,c_i} \ge \left( (1+\eps) (RN)^s
 \Bigl( \sum_{i=1}^N \sum_{z \in\Sigma} w_{i,z}^{s+1}\Bigr) \right)^{1/(s+1)} \
  . \]
For $s=1$, this soft decoding condition is identical to the one for
  Reed-Solomon codes in \cite{GS98}.
\end{remark}
 
\subsection{Binary codes decodable up to Zyablov bound}

Concatenating the folded RS codes with suitable inner codes also gives
us polytime constructible binary codes that can be efficiently list
decoded up to the Zyablov bound, i.e., up to twice the radius achieved
by the standard GMD decoding of concatenated codes.  The optimal list
recoverability of the folded RS codes plays a crucial role in
establishing such a result.
\begin{theorem}
\label{thm:binary-ld-zyablov}
For all $0 < R,r < 1$ and all $\eps > 0$, there is a polynomial time
constructible family of binary linear codes of rate at least $R\cdot
r$ which can be list decoded in polynomial time up to a fraction
$(1-R) H^{-1}(1-r) - \eps$ of errors.
\end{theorem}
\begin{proof}
We will construct binary codes with the
claimed property by concatenating two codes $C_1$ and $C_2$. For
$C_1$, we will use a folded RS code over a field of characteristic $2$
with block length $n_1$,
rate at least $R$, and which can be
$(R+\eps,l)$-list recovered in polynomial time for $l = \lceil 10/\eps
\rceil$. Let the 
alphabet size of $C_1$ be $2^M$ where $M = O(\eps^{-2}\log(1/\eps) \log n_1)$. 
For $C_2$, we will use a binary linear code of dimension $M$ and rate at
least $r$ which is $(\p,l)$-list decodable for $\p =
H^{-1}(1-r-\eps)$. Such a code is known to exist via a random coding
argument that employs the semi-random method~\cite{GHSZ}. Also, a greedy
construction of such a code by constructing its $M$ basis elements in
turn is presented in \cite{GHSZ} and this process takes
$2^{O(M)}$ time. We conclude that the necessary inner code can be
constructed in $n_1^{O(\eps^{-2}\log(1/\eps))}$ time. The code $C_1$, being a folded
RS code over a field of characteristic $2$, is $\F_2$-linear, and
therefore when concatenated with a binary linear inner code such as
$C_2$, results in a binary linear code. The rate of the concatenated
code is at least $R \cdot r$.

The decoding algorithm proceeds in a natural way. Given a received word,
we break it up into blocks corresponding to the various inner
encodings by $C_1$. Each of these blocks is list decoded up to a
radius $\p$, returning a set of at most $l$ possible candidates for
each outer codeword symbol. The outer code is then $(R+\eps,l)$-list
recovered using these sets, each of which has size at most $l$, as input. 
To argue about the fraction of errors this algorithm corrects, we note
that the algorithm fails to recover a codeword only if on more than a
fraction $(1-R-\eps)$ of the inner blocks the codeword differs from
the received word on more than a fraction $\p$ of symbols. It follows
that the algorithm correctly list decodes up to a radius $(1-R-\eps) \p
= (1-R-\eps) H^{-1}(1-r-\eps)$. Since $\eps > 0$ was arbitrary, we get
the claimed result.
\end{proof}

Optimizing over the choice of inner and outer codes rates $r,R$ in the
above results, we can decode up to the Zyablov bound, see Figure
\ref{fig:binary-ld}. 

\begin{remark}
  In particular, decoding up to the Zyablov bound implies that we can
  correct a fraction $(1/2-\eps)$ of errors with rate $\Omega(\eps^3)$
  for small $\eps \to 0$, which is better than the rate of
  $\Omega(\eps^3/\log (1/\eps))$ achieved in \cite{GP-focs06}.
  However, our construction and decoding complexity are
  $n^{O(\eps^{-2}\log(1/\eps))}$ whereas these are at most $f(\eps)
  n^c$ for an absolute constant $c$ in \cite{GP-focs06}.  Also, we
  bound the list size needed in the worst-case by
  $n^{O(\eps^{-1}\log(1/\eps))}$, while the list size needed in the
  construction in \cite{GP-focs06} is $(1/\eps)^{O(\log \log
    (1/\eps))}$.
\end{remark}

\begin{remark}[Decoding up to the Blokh-Zyablov bound]
  In a follow-up paper, we use a similar approach extended to
  multilevel concatenation schemes together with inner codes that have
  good ``nested'' list-decodability properties, to construct binary
  codes list-decodable up to the {\em Blokh-Zyablov} bound~\cite{GR-random07}.
\end{remark}

\subsection{Capacity-Achieving codes over smaller alphabets}
\label{sec:const-alphabets}

Our result of Theorem~\ref{thm:final-capacity} has two undesirable
aspects: both the alphabet size and worst-case list size output by the
list decoding algorithm are a polynomial of large degree in the block
length. We now show that the alphabet size can be reduced to a
constant that depends only on the distance $\eps$ to capacity.


\begin{theorem}
\label{thm:cap-const-alph}
For every $R$, $0 < R < 1$, every $\eps > 0$, there is a polynomial
time constructible family of codes over an alphabet of size
$2^{O(\eps^{-4} \log (1/\eps))}$ that have rate at least $R$ and which
can be list decoded up to a fraction $(1-R-\epsilon)$ of errors in
polynomial time.
\end{theorem}
\begin{proof} 
  The theorem is proved using the code construction scheme used in
  \cite{GI-ieeejl} for linear time unique decodable codes with optimal
  rate, with different components appropriate for list decoding
  plugged in. We briefly describe the main ideas behind the
  construction and proof below. The high level approach is to
  concatenate two codes $C_{\rm out}$ and $C_{\rm in}$, and then
  redistribute the symbols of the resulting codeword using an expander
  graph (Figure~\ref{fig:capacity-small} depicts this high level
  structure and should be useful in reading the following formal
  description). In the following, assume that $\eps < 1/6$ and let
  $\delta = \eps^2$.

The outer code $C_{\rm out}$ will be a code of rate $(1-2\eps)$ over
an alphabet $\Sigma$ of size $n^{(1/\delta)^{O(1)}}$ that can be
$(1-\eps, O(1/\eps))$-list recovered in polynomial time, as
guaranteed by Theorem~\ref{thm:main-LR}. That is, the rate of $C_{\rm
out}$ will be close to $1$, and it can be $(\zeta,l)$-list recovered
for large $l$ and $\zeta \to 1$.

The inner code $C_{\rm in}$ will be a $((1-R-4\eps),
O(1/\eps))$-list decodable code with near-optimal rate, say rate at
least $(R+3\eps)$. Such a code is guaranteed to exist over an
alphabet of size $O(1/\eps^2)$ using random coding arguments. A
naive brute-force for such a code, however, is too expensive, since we
need a code with $|\Sigma| = n^{\Omega(1)}$ codewords. Guruswami and
Indyk~\cite{GI-focs01}, see also \cite[Sec.  9.3]{G-thesis}, prove
that there is a small (quasi-polynomial sized) sample space of {\em
  pseudolinear codes} in which most codes have the needed property.
Furthermore, they also present a deterministic polynomial time
construction of such a code (using derandomization techniques), see
\cite[Sec. 9.3.3]{G-thesis}.

The concatenation of $C_{\rm out}$ and $C_{\rm in}$ gives a code
$C_{\rm concat}$ of rate at least $(1-2\eps)(R+3\eps) \ge R$ over an
alphabet $\Sigma$ of size $|\Sigma|=O(1/\eps^2)$.  Moreover, given a
received word of the concatenated code, one can find all codewords
that agree with the received word on a fraction $R+4\eps$ of locations
in at least $(1-\eps)$ of the inner blocks. Indeed, we can do this by
running the natural list decoding algorithm, call it ${\cal A}$, for
$C_{\rm concat}$ that decodes each of the inner blocks to a radius of
$(1-R-4\eps)$ returning up to $l = O(1/\eps)$ possibilities for each
block, and then $(1-\eps,l)$-list recovering $C_{\rm out}$.

The last component in this construction is a $D =
O(1/\eps^4)$-regular bipartite expander graph which is used to redistribute
symbols of the concatenated code in a manner so that an overall
agreement on a fraction $R+7\eps$ of the redistributed symbols
implies a fractional agreement of at least $R+4\eps$ on most
(specifically a fraction $(1-\eps)$) of the inner blocks of the
concatenated code. In other words, the expander redistributes symbols
in a manner that ``smoothens'' the distributions of errors evenly
among the various inner blocks (except for possibly a $\eps$
fraction of the blocks).  This expander based redistribution incurs no
loss in rate, but increases the alphabet size to
$O(1/\eps^2)^{O(1/\eps^4)} = 2^{O(\eps^{-4} \log(1/\eps))}$.

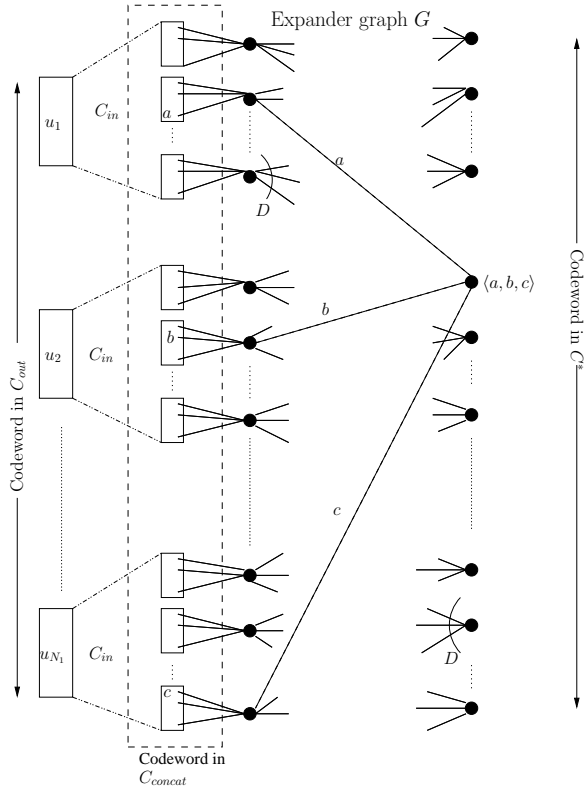
\begin{figure}[ht]
\begin{center}
\resizebox{!}{4.1in}{\input{capacity.pstex_t}}
\caption[Capacity Achieving Code over Small Alphabet]{The code $C^*$ used
in the proof of Theorem~\ref{thm:cap-const-alph}. We start with a codeword
$\angles{u_1,\dots,u_{N_1}}$ in $C_{\rm out}$. Then every symbol is encoded
by $C_{\rm in}$ to form a codeword in $C_{\rm concat}$ (this intermediate codeword is
marked by the dotted box). The symbols in the codeword for $C_{\rm concat}$ are
divided into chunks of $D$ symbols and then redistributed along the edges of
an expander $G$ of degree $D$.
In the figure, we use $D=3$ for clarity. Also the distribution
of three symbols $a$, $b$ and $c$ (that form a symbol in the final codeword
in $C^*$) is shown.}
\label{fig:capacity-small}
\end{center}
\end{figure}

We now discuss some details of how the expander is used.  Suppose that
the block length of the folded RS code $C_{\rm out}$ is $N_1$ and that
of $C_{\rm in}$ is $N_2$. Let us assume that $N_2$ is a multiple of
$D$, say $N_2 = n_2 D$ (if this is not the case, we can make it so by
padding at most $D-1$ dummy symbols at a negligible loss in
rate). Therefore codewords of $C_{\rm in}$, and therefore also of
$C_{\rm concat}$, can be thought of as being composed of blocks of $D$
symbols each. Let $N = N_1 n_2$, so that codewords of $C_{\rm concat}$
can be viewed as elements in $(\Sigma^D)^N$.

Let $G=(L,R,E)$ be a $D$-regular
bipartite graph with $N$ vertices on each side (i.e., $|L| = |R| =
N$), with the property that for every subset $Y \subseteq R$ of size
at least $(R+7\eps)N$, the number of vertices belonging to $L$ that
have at most $(R+6\eps)D$ of their neighbors in $Y$ is at most $\delta
N$ (for $\delta = \eps^2$). It is a well-known fact (used also in
\cite{GI-ieeejl}) that if $G$ is picked to be the double cover of a
Ramanujan expander of degree $D \ge 4/(\delta \eps^2)$, then $G$ will
have such a property.

We now define our final code $C^* = G(C_{\rm concat}) \subseteq
(\Sigma^D)^N$ formally. The codewords in $C^*$ are in one-one
correspondence with those of $C_{\rm concat}$. Given a codeword $c \in
C_{\rm concat}$, its $N D$ symbols (each belonging to $\Sigma$) are
placed on the $N D$ edges of $G$, with the $D$ symbols in its $i$'th
block (belonging to $\Sigma^D$, as defined above) being placed on the
$D$ edges incident on the $i$'th vertex of $L$ (in some fixed order).
The codeword in $C^*$ corresponding to $c$ has as its $i$'th symbol
the collection of $D$ symbols (in some fixed order) on the $D$ edges
incident on the $i$'th vertex of $R$. See Figure~\ref{fig:capacity-small}
for a pictorial view of the construction.

Note that the rate of $C^*$ is identical to that $C_{\rm concat}$, and
is thus at least $R$. Its alphabet size is $|\Sigma|^D =
O(1/\eps^2)^{O(1/\eps^4)} = 2^{O(\eps^{-4} \log(1/\eps))}$, as
claimed. We will now argue how $C^*$ can be list decoded up to
a fraction $(1-R-7\eps)$ of errors.

Given a received word $\mv{r} \in (\Sigma^D)^N$, the following is the
natural algorithm to find all codewords of $C^*$ with agreement at
least $(R+7\eps) N$ with $\mv{r}$. Redistribute symbols according to
the expander backwards to compute the received word $\mv{r'}$ for $C_{\rm
concat}$ which would result in $\mv{r}$. Then run the earlier-mentioned decoding algorithm ${\cal A}$ on $\mv{r'}$.

We now briefly argue the correctness of this algorithm. Let $\mv{c}
\in C^*$ be a codeword with agreement at least $(R+7\eps) N$ with
$\mv{r}$. Let $\mv{c'}$ denote the codeword of $C_{\rm concat}$ that
leads to $\mv{c}$ after symbol redistribution by $G$, and finally
suppose $\mv{c''}$ is the codeword of $C_{\rm out}$ that yields
$\mv{c'}$ upon concatenation by $C_{\rm in}$. By the expansion
properties of $G$, it follows that all but a $\delta$ fraction of $N$
$D$-long blocks of $\mv{r'}$ have agreement at least $(R+6\eps) D$ with the
corresponding blocks of $\mv{c'}$. By an averaging argument, this
implies that at least a fraction $(1-\sqrt{\delta})$ of the $N_1$
blocks of $\mv{c'}$ that correspond to codewords of $C_{\rm in}$
encoding the $N_1$ symbols of $\mv{c''}$, agree with at least a
fraction $(1-\sqrt{\delta})(R+6\eps)=(1-\eps)(R+6\eps) \ge R+4\eps$ of
the symbols of the corresponding block of $\mv{r'}$.  As argued
earlier, this in turn implies that the decoding algorithm ${\cal A}$
for $C_{\rm concat}$ when run on input $\mv{r'}$ will output a
polynomial size list that will include $\mv{c'}$.
\end{proof}

\section{Concluding Remarks}
\label{sec:concl}

We close with some remarks and open questions.  In the preliminary
version~\cite{GR-stoc06} of this paper, we noted that the folded RS
codes bear some resemblance to certain ``randomness extractors''
constructed in \cite{SU}, and wondered if some of the techniques in
this work and \cite{PV-focs05} could be used to construct simple
extractors based on univariate polynomials. In a recent
work~\cite{GUV}, this has been answered in the affirmative in a fairly
strong sense. It is shown in \cite{GUV} that the Parvaresh-Vardy codes
yield excellent ``randomness condensers,'' which achieve near-optimal
compression of a weak random source while preserving all its
min-entropy, and in turn these lead to the best known randomness
extractors (that are optimal up to constant factors).

We have solved the qualitative problem of achieving list decoding
capacity over large alphabets.  Our work could be improved with some
respect to some parameters.  The size of the list needed to perform
list decoding to a radius that is within $\eps$ of capacity grows as
$n^{O(1/\eps)}$ where $n$ is the block length of the code. It remains
an open question to bring this list size down to a constant
independent of $n$, or even to $f(\eps) n^{c}$ with an exponent $c$
independent of $\eps$ (we recall that the existential random coding
arguments work with a list size of $O(1/\eps)$).  We managed to reduce
the alphabet size needed to approach capacity to a constant
independent of $n$. However, this involved a brute-force search for a
rather large code.  Obtaining a ``direct'' algebraic construction over
a constant-sized alphabet (such as variants of algebraic-geometric (AG)
codes) might help in addressing these two issues. To this end,
Guruswami and Patthak~\cite{GP-focs06} define {\em correlated AG
  codes}, and describe list decoding algorithms for those codes, based
on a generalization of the Parvaresh-Vardy approach to the general
class of algebraic-geometric codes (of which RS codes are a special
case). However, to relate folded AG codes to correlated AG codes like
we did for RS codes requires bijections on the set of rational points
of the underlying algebraic curve that have some special, hard to
guarantee, property.  This step seems like an highly intricate
algebraic task, and especially so in the interesting asymptotic
setting of a family of asymptotically good AG codes over a fixed
alphabet.

Finally, constructing binary codes (or $q$-ary codes for some fixed,
small value of $q$) that approach the respective list decoding
capacity remains a challenging open problem. In recent
work~\cite{GR-soda08}, we show that there {\em exist} $q$-ary linear
concatenated codes that achieve list decoding capacity (in the sense
that every Hamming ball of radius $H_q^{-1}(1-R-\eps)$ has
polynomially many codewords, where $R$ is the rate). In particular,
this results holds when the outer code is a folded RS code.  This is
somewhat encouraging news since concatenation has been the preeminent
method to construct good list-decodable codes over small
alphabets. But realizing the full potential of concatenated codes and
achieving capacity (or even substantially improving upon the
Blokh-Zyablov bound) with explicit codes and polynomial time decoding
remains a huge challenge. It seems likely that carefully chosen soft
information to pass from the inner decodings to the outer algebraic
decoder (see \cite{koetter-itw06,GS-soft-02} for examples of such
decoders) may hold the key to further progress in list decoding
concatenated codes.

\section*{Acknowledgments}
We thank J\o rn Justesen for suggesting the alternate bound on
decoding radius (\ref{eq:high-rate}) discussed in
Section~\ref{sec:tri-prac}, and for kindly allowing us to include it
in our presentation. We thank an anonymous referee for several useful
comments and in particular for encouraging us to highlight the
relevance of the results even for small values of the folding
parameter; this was the impetus for the discussion in
Section~\ref{sec:tri-prac}. We thank Piotr Indyk, Charanjit Jutla,
Farzad Parvaresh, Anindya Patthak, Madhu Sudan, and Alexander Vardy
for useful discussions and comments.

\bibliographystyle{abbrv}
\bibliography{list-decoding}

\end{document}

%% file: frs.pstex_t
\begin{picture}(0,0)%
\includegraphics{frs.pstex}%
\end{picture}%
\setlength{\unitlength}{3947sp}%
\begingroup\makeatletter\ifx\SetFigFont\undefined%
\gdef\SetFigFont#1#2#3#4#5{%
  \reset@font\fontsize{#1}{#2pt}%
  \fontfamily{#3}\fontseries{#4}\fontshape{#5}%
  \selectfont}%
\fi\endgroup%
\begin{picture}(12865,4266)(289,-4894)
\put(376,-1036){\makebox(0,0)[lb]{\smash{{\SetFigFont{14}{16.8}{\rmdefault}{\mddefault}{\updefault}{\color[rgb]{0,0,0}$f(x_0)$}%
}}}}
\put(1276,-1036){\makebox(0,0)[lb]{\smash{{\SetFigFont{14}{16.8}{\rmdefault}{\mddefault}{\updefault}{\color[rgb]{0,0,0}$f(x_1)$}%
}}}}
\put(2176,-1036){\makebox(0,0)[lb]{\smash{{\SetFigFont{14}{16.8}{\rmdefault}{\mddefault}{\updefault}{\color[rgb]{0,0,0}$f(x_2)$}%
}}}}
\put(3076,-1036){\makebox(0,0)[lb]{\smash{{\SetFigFont{14}{16.8}{\rmdefault}{\mddefault}{\updefault}{\color[rgb]{0,0,0}$f(x_3)$}%
}}}}
\put(3976,-1036){\makebox(0,0)[lb]{\smash{{\SetFigFont{14}{16.8}{\rmdefault}{\mddefault}{\updefault}{\color[rgb]{0,0,0}$f(x_4)$}%
}}}}
\put(4876,-1036){\makebox(0,0)[lb]{\smash{{\SetFigFont{14}{16.8}{\rmdefault}{\mddefault}{\updefault}{\color[rgb]{0,0,0}$f(x_5)$}%
}}}}
\put(5776,-1036){\makebox(0,0)[lb]{\smash{{\SetFigFont{14}{16.8}{\rmdefault}{\mddefault}{\updefault}{\color[rgb]{0,0,0}$f(x_6)$}%
}}}}
\put(6676,-1036){\makebox(0,0)[lb]{\smash{{\SetFigFont{14}{16.8}{\rmdefault}{\mddefault}{\updefault}{\color[rgb]{0,0,0}$f(x_7)$}%
}}}}
\put(9301,-1036){\makebox(0,0)[lb]{\smash{{\SetFigFont{14}{16.8}{\rmdefault}{\mddefault}{\updefault}{\color[rgb]{0,0,0}$f(x_{n-4})$}%
}}}}
\put(10201,-1036){\makebox(0,0)[lb]{\smash{{\SetFigFont{14}{16.8}{\rmdefault}{\mddefault}{\updefault}{\color[rgb]{0,0,0}$f(x_{n-3})$}%
}}}}
\put(11101,-1036){\makebox(0,0)[lb]{\smash{{\SetFigFont{14}{16.8}{\rmdefault}{\mddefault}{\updefault}{\color[rgb]{0,0,0}$f(x_{n-2})$}%
}}}}
\put(12001,-1036){\makebox(0,0)[lb]{\smash{{\SetFigFont{14}{16.8}{\rmdefault}{\mddefault}{\updefault}{\color[rgb]{0,0,0}$f(x_{n-1})$}%
}}}}
\put(4951,-2836){\makebox(0,0)[lb]{\smash{{\SetFigFont{14}{16.8}{\rmdefault}{\mddefault}{\updefault}{\color[rgb]{0,0,0}$f(x_0)$}%
}}}}
\put(4951,-3436){\makebox(0,0)[lb]{\smash{{\SetFigFont{14}{16.8}{\rmdefault}{\mddefault}{\updefault}{\color[rgb]{0,0,0}$f(x_1)$}%
}}}}
\put(4951,-4036){\makebox(0,0)[lb]{\smash{{\SetFigFont{14}{16.8}{\rmdefault}{\mddefault}{\updefault}{\color[rgb]{0,0,0}$f(x_2)$}%
}}}}
\put(4951,-4636){\makebox(0,0)[lb]{\smash{{\SetFigFont{14}{16.8}{\rmdefault}{\mddefault}{\updefault}{\color[rgb]{0,0,0}$f(x_3)$}%
}}}}
\put(5851,-2836){\makebox(0,0)[lb]{\smash{{\SetFigFont{14}{16.8}{\rmdefault}{\mddefault}{\updefault}{\color[rgb]{0,0,0}$f(x_4)$}%
}}}}
\put(5851,-3436){\makebox(0,0)[lb]{\smash{{\SetFigFont{14}{16.8}{\rmdefault}{\mddefault}{\updefault}{\color[rgb]{0,0,0}$f(x_5)$}%
}}}}
\put(5851,-4036){\makebox(0,0)[lb]{\smash{{\SetFigFont{14}{16.8}{\rmdefault}{\mddefault}{\updefault}{\color[rgb]{0,0,0}$f(x_6)$}%
}}}}
\put(5851,-4636){\makebox(0,0)[lb]{\smash{{\SetFigFont{14}{16.8}{\rmdefault}{\mddefault}{\updefault}{\color[rgb]{0,0,0}$f(x_7)$}%
}}}}
\put(7876,-2836){\makebox(0,0)[lb]{\smash{{\SetFigFont{14}{16.8}{\rmdefault}{\mddefault}{\updefault}{\color[rgb]{0,0,0}$f(x_{n-4})$}%
}}}}
\put(7876,-3436){\makebox(0,0)[lb]{\smash{{\SetFigFont{14}{16.8}{\rmdefault}{\mddefault}{\updefault}{\color[rgb]{0,0,0}$f(x_{n-3})$}%
}}}}
\put(7876,-4036){\makebox(0,0)[lb]{\smash{{\SetFigFont{14}{16.8}{\rmdefault}{\mddefault}{\updefault}{\color[rgb]{0,0,0}$f(x_{n-2})$}%
}}}}
\put(7876,-4636){\makebox(0,0)[lb]{\smash{{\SetFigFont{14}{16.8}{\rmdefault}{\mddefault}{\updefault}{\color[rgb]{0,0,0}$f(x_{n-1})$}%
}}}}
\end{picture}%

%% file: reduction-journal.pstex_t
\begin{picture}(0,0)%
\includegraphics{reduction-journal.pstex}%
\end{picture}%
\setlength{\unitlength}{3947sp}%
\begingroup\makeatletter\ifx\SetFigFont\undefined%
\gdef\SetFigFont#1#2#3#4#5{%
  \reset@font\fontsize{#1}{#2pt}%
  \fontfamily{#3}\fontseries{#4}\fontshape{#5}%
  \selectfont}%
\fi\endgroup%
\begin{picture}(13106,5191)(128,-4769)
\put(6451,-2536){\makebox(0,0)[lb]{\smash{{\SetFigFont{12}{14.4}{\rmdefault}{\mddefault}{\updefault}{\color[rgb]{0,0,0}$\mathrm{FRS}$ codeword}%
}}}}
\put(3226,-61){\makebox(0,0)[lb]{\smash{{\SetFigFont{14}{16.8}{\rmdefault}{\mddefault}{\updefault}{\color[rgb]{0,0,0}$f(x_0)$}%
}}}}
\put(3076,-661){\makebox(0,0)[lb]{\smash{{\SetFigFont{14}{16.8}{\rmdefault}{\mddefault}{\updefault}{\color[rgb]{0,0,0}$f(\gamma x_0)$}%
}}}}
\put(3076,-1261){\makebox(0,0)[lb]{\smash{{\SetFigFont{14}{16.8}{\rmdefault}{\mddefault}{\updefault}{\color[rgb]{0,0,0}$f(\gamma^2 x_0)$}%
}}}}
\put(5251,-61){\makebox(0,0)[lb]{\smash{{\SetFigFont{14}{16.8}{\rmdefault}{\mddefault}{\updefault}{\color[rgb]{0,0,0}$f(x_0)$}%
}}}}
\put(5251,-661){\makebox(0,0)[lb]{\smash{{\SetFigFont{14}{16.8}{\rmdefault}{\mddefault}{\updefault}{\color[rgb]{0,0,0}$f(\gamma x_0)$}%
}}}}
\put(5251,-1261){\makebox(0,0)[lb]{\smash{{\SetFigFont{14}{16.8}{\rmdefault}{\mddefault}{\updefault}{\color[rgb]{0,0,0}$f(\gamma^2 x_0)$}%
}}}}
\put(5251,-1861){\makebox(0,0)[lb]{\smash{{\SetFigFont{14}{16.8}{\rmdefault}{\mddefault}{\updefault}{\color[rgb]{0,0,0}$f(\gamma^3 x_0)$}%
}}}}
\put(6226,-61){\makebox(0,0)[lb]{\smash{{\SetFigFont{14}{16.8}{\rmdefault}{\mddefault}{\updefault}{\color[rgb]{0,0,0}$f(x_4)$}%
}}}}
\put(6226,-661){\makebox(0,0)[lb]{\smash{{\SetFigFont{14}{16.8}{\rmdefault}{\mddefault}{\updefault}{\color[rgb]{0,0,0}$f(\gamma x_4)$}%
}}}}
\put(6226,-1261){\makebox(0,0)[lb]{\smash{{\SetFigFont{14}{16.8}{\rmdefault}{\mddefault}{\updefault}{\color[rgb]{0,0,0}$f(\gamma^2 x_4)$}%
}}}}
\put(6226,-1861){\makebox(0,0)[lb]{\smash{{\SetFigFont{14}{16.8}{\rmdefault}{\mddefault}{\updefault}{\color[rgb]{0,0,0}$f(\gamma^3 x_4)$}%
}}}}
\put(676,-3436){\makebox(0,0)[lb]{\smash{{\SetFigFont{14}{16.8}{\rmdefault}{\mddefault}{\updefault}{\color[rgb]{0,0,0}$f(x_0)$}%
}}}}
\put(676,-4036){\makebox(0,0)[lb]{\smash{{\SetFigFont{14}{16.8}{\rmdefault}{\mddefault}{\updefault}{\color[rgb]{0,0,0}$f(\gamma x_0)$}%
}}}}
\put(1651,-3436){\makebox(0,0)[lb]{\smash{{\SetFigFont{14}{16.8}{\rmdefault}{\mddefault}{\updefault}{\color[rgb]{0,0,0}$f(\gamma x_0)$}%
}}}}
\put(1651,-4036){\makebox(0,0)[lb]{\smash{{\SetFigFont{14}{16.8}{\rmdefault}{\mddefault}{\updefault}{\color[rgb]{0,0,0}$f(\gamma^2 x_0)$}%
}}}}
\put(2626,-3436){\makebox(0,0)[lb]{\smash{{\SetFigFont{14}{16.8}{\rmdefault}{\mddefault}{\updefault}{\color[rgb]{0,0,0}$f(\gamma^2 x_0)$}%
}}}}
\put(2626,-4036){\makebox(0,0)[lb]{\smash{{\SetFigFont{14}{16.8}{\rmdefault}{\mddefault}{\updefault}{\color[rgb]{0,0,0}$f(\gamma^3 x_0)$}%
}}}}
\put(3151,-1861){\makebox(0,0)[lb]{\smash{{\SetFigFont{14}{16.8}{\rmdefault}{\mddefault}{\updefault}{\color[rgb]{0,0,0}$f(\gamma^3 x_0)$}%
}}}}
\put(5776,-4711){\makebox(0,0)[lb]{\smash{{\SetFigFont{12}{14.4}{\rmdefault}{\mddefault}{\updefault}{\color[rgb]{0,0,0}$\mathrm{PV}$ codeword}%
}}}}
\put(3601,-3436){\makebox(0,0)[lb]{\smash{{\SetFigFont{14}{16.8}{\rmdefault}{\mddefault}{\updefault}{\color[rgb]{0,0,0}$f(x_4)$}%
}}}}
\put(3526,-4036){\makebox(0,0)[lb]{\smash{{\SetFigFont{14}{16.8}{\rmdefault}{\mddefault}{\updefault}{\color[rgb]{0,0,0}$f(\gamma x_4)$}%
}}}}
\put(4501,-3436){\makebox(0,0)[lb]{\smash{{\SetFigFont{14}{16.8}{\rmdefault}{\mddefault}{\updefault}{\color[rgb]{0,0,0}$f(\gamma x_4)$}%
}}}}
\put(4576,-4036){\makebox(0,0)[lb]{\smash{{\SetFigFont{14}{16.8}{\rmdefault}{\mddefault}{\updefault}{\color[rgb]{0,0,0}$f(\gamma^2 x_4)$}%
}}}}
\put(5476,-3436){\makebox(0,0)[lb]{\smash{{\SetFigFont{14}{16.8}{\rmdefault}{\mddefault}{\updefault}{\color[rgb]{0,0,0}$f(\gamma^2 x_4)$}%
}}}}
\put(5476,-4036){\makebox(0,0)[lb]{\smash{{\SetFigFont{14}{16.8}{\rmdefault}{\mddefault}{\updefault}{\color[rgb]{0,0,0}$f(\gamma^3 x_4)$}%
}}}}
\end{picture}%

%% file: capacity.pstex_t
\begin{picture}(0,0)%
\includegraphics{capacity.pstex}%
\end{picture}%
\setlength{\unitlength}{3947sp}%
\begingroup\makeatletter\ifx\SetFigFont\undefined%
\gdef\SetFigFont#1#2#3#4#5{%
  \reset@font\fontsize{#1}{#2pt}%
  \fontfamily{#3}\fontseries{#4}\fontshape{#5}%
  \selectfont}%
\fi\endgroup%
\begin{picture}(7738,10605)(957,-9829)
\put(3076,-3811){\makebox(0,0)[lb]{\smash{{\SetFigFont{14}{16.8}{\rmdefault}{\mddefault}{\updefault}{\color[rgb]{0,0,0}$b$}%
}}}}
\put(5326,-6136){\makebox(0,0)[lb]{\smash{{\SetFigFont{14}{16.8}{\rmdefault}{\mddefault}{\updefault}{\color[rgb]{0,0,0}$c$}%
}}}}
\put(5176,-3436){\makebox(0,0)[lb]{\smash{{\SetFigFont{14}{16.8}{\rmdefault}{\mddefault}{\updefault}{\color[rgb]{0,0,0}$b$}%
}}}}
\put(7351,-3061){\makebox(0,0)[lb]{\smash{{\SetFigFont{14}{16.8}{\rmdefault}{\mddefault}{\updefault}{\color[rgb]{0,0,0}$\langle a,b,c\rangle$}%
}}}}
\put(4276,-2086){\makebox(0,0)[lb]{\smash{{\SetFigFont{14}{16.8}{\rmdefault}{\mddefault}{\updefault}{\color[rgb]{0,0,0}$D$}%
}}}}
\put(6826,-8161){\makebox(0,0)[lb]{\smash{{\SetFigFont{14}{16.8}{\rmdefault}{\mddefault}{\updefault}{\color[rgb]{0,0,0}$D$}%
}}}}
\put(2101,-736){\makebox(0,0)[lb]{\smash{{\SetFigFont{14}{16.8}{\rmdefault}{\mddefault}{\updefault}{\color[rgb]{0,0,0}$C_{in}$}%
}}}}
\put(2026,-4036){\makebox(0,0)[lb]{\smash{{\SetFigFont{14}{16.8}{\rmdefault}{\mddefault}{\updefault}{\color[rgb]{0,0,0}$C_{in}$}%
}}}}
\put(2026,-8086){\makebox(0,0)[lb]{\smash{{\SetFigFont{14}{16.8}{\rmdefault}{\mddefault}{\updefault}{\color[rgb]{0,0,0}$C_{in}$}%
}}}}
\put(1126,-5836){\rotatebox{90.0}{\makebox(0,0)[lb]{\smash{{\SetFigFont{14}{16.8}{\rmdefault}{\mddefault}{\updefault}{\color[rgb]{0,0,0}Codeword in $C_{out}$}%
}}}}}
\put(8551,-2536){\rotatebox{270.0}{\makebox(0,0)[lb]{\smash{{\SetFigFont{14}{16.8}{\rmdefault}{\mddefault}{\updefault}{\color[rgb]{0,0,0}Codeword in $C^*$}%
}}}}}
\put(3016,-766){\makebox(0,0)[lb]{\smash{{\SetFigFont{14}{16.8}{\rmdefault}{\mddefault}{\updefault}{\color[rgb]{0,0,0}$a$}%
}}}}
\put(3031,-8611){\makebox(0,0)[lb]{\smash{{\SetFigFont{14}{16.8}{\rmdefault}{\mddefault}{\updefault}{\color[rgb]{0,0,0}$c$}%
}}}}
\put(5356,-1396){\makebox(0,0)[lb]{\smash{{\SetFigFont{14}{16.8}{\rmdefault}{\mddefault}{\updefault}{\color[rgb]{0,0,0}$a$}%
}}}}
\put(4486,479){\makebox(0,0)[lb]{\smash{{\SetFigFont{17}{20.4}{\rmdefault}{\mddefault}{\updefault}{\color[rgb]{0,0,0}Expander graph $G$}%
}}}}
\put(1426,-886){\makebox(0,0)[lb]{\smash{{\SetFigFont{14}{16.8}{\rmdefault}{\mddefault}{\updefault}{\color[rgb]{0,0,0}$u_1$}%
}}}}
\put(1426,-4036){\makebox(0,0)[lb]{\smash{{\SetFigFont{14}{16.8}{\rmdefault}{\mddefault}{\updefault}{\color[rgb]{0,0,0}$u_2$}%
}}}}
\put(1351,-8086){\makebox(0,0)[lb]{\smash{{\SetFigFont{14}{16.8}{\rmdefault}{\mddefault}{\updefault}{\color[rgb]{0,0,0}$u_{N_1}$}%
}}}}
\put(2701,-9766){\makebox(0,0)[lb]{\smash{{\SetFigFont{14}{16.8}{\rmdefault}{\mddefault}{\updefault}{\color[rgb]{0,0,0}$C_{concat}$}%
}}}}
\end{picture}%